\documentclass{WileyMSP-template}
\usepackage{ragged2e}
\AtBeginDocument{\justifying}
\usepackage{indentfirst}
\usepackage{graphicx}
\usepackage[space]{grffile}
\usepackage{latexsym}
\usepackage{textcomp}
\usepackage{longtable}
\usepackage{tabulary}
\usepackage{booktabs,array,multirow}
\usepackage{amsfonts,amsmath,amssymb}
\usepackage{natbib}
\usepackage{url}
\usepackage{hyperref}
\hypersetup{colorlinks=false,pdfborder={0 0 0}}
\usepackage{etoolbox}
\usepackage{amsmath} 
\newcommand{\angstrom}{\text{\normalfont\AA}}
\newif\iflatexml\latexmlfalse
\usepackage{bm}
\usepackage{algorithm}
\usepackage{algpseudocode}
\AtBeginDocument{\DeclareGraphicsExtensions{.pdf,.PDF,.eps,.EPS,.png,.PNG,.tif,.TIF,.jpg,.JPG,.jpeg,.JPEG}}
\usepackage[utf8]{inputenc}
\usepackage[english]{babel}

\usepackage{siunitx}

\iflatexml

\else

\usepackage{lineno}

\begin{document}
	
	\pagestyle{fancy}
	
	\title{Gap-free Information Transfer in 4D-STEM via Fusion of Complementary Scattering Channels}
	
	\maketitle
	
	\author{Shengbo You}
	\author{Georgios Varnavides}
	\author{Sagar Khavnekar}
	\author{Nikita Palatkin}
	\author{Sihan Shao}
	\author{Mingjian Wu}
	\author{Daniel Stroppa}
	\author{Darya Chernikova}
	\author{Baixu Zhu}
	\author{Ricardo Egoavil}
	\author{Stefano Vespucci}
	\author{Xingchen Ye}
	\author{Florian K.\,M. Schur}
	\author{Erdmann Spiecker}
	\author{Philipp Pelz*}
	
	\dedication{}
	
	\begin{affiliations}
		S. You\\
		Institute of Micro- and Nanostructure Research (IMN) \& Center for Nanoanalysis and Electron Microscopy (CENEM),\\
		Friedrich Alexander-Universität Erlangen-Nürnberg, IZNF, 91058 Erlangen, Germany\\
		
		\vspace{0.2cm}
		G. Varnavides\\
		Department of Imaging Physics, Delft University of Technology,\\
		Lorentzweg 1, 2628 CJ Delft, The Netherlands\\
		
		\vspace{0.2cm}
		S. Khavnekar, R. Egoavil, S. Vespucci\\
		Materials and Structural Analysis Division, Thermo Fisher Scientific,\\
		Achtseweg Noord, 5651 GG Eindhoven, The Netherlands\\
		
		\vspace{0.2cm}
		D. Stroppa\\
		DSTL GmbH, Baden, Switzerland\\
		
		\vspace{0.2cm}
		D. Chernikova, F. K.\,M. Schur\\
		Institute of Science and Technology Austria (ISTA),\\
		3400 Klosterneuburg, Austria\\
		
		\vspace{0.2cm}
		B. Zhu, X. Ye\\
		Department of Chemistry, Indiana University,\\
		800 E. Kirkwood Avenue, Bloomington, Indiana 47405, United States\\
		
		\vspace{0.2cm}
		N. Palatkin, S. Shao, M. Wu, E. Spiecker. P.M. Pelz\\
		Institute of Micro- and Nanostructure Research (IMN) \& Center for Nanoanalysis and Electron Microscopy (CENEM),\\
		Friedrich Alexander-Universität Erlangen-Nürnberg, IZNF, 91058 Erlangen, Germany\\
		Email: philipp.pelz@fau.de\\

	\end{affiliations}
	
	
	\begin{abstract}
		Linear phase-contrast scanning transmission electron microscopy (STEM) techniques compatible with high-throughput 4D-STEM acquisition are widely used to enhance phase contrast in weakly scattering and beam-sensitive materials. In these modalities, contrast transfer is often suppressed at low spatial frequencies, resulting in a characteristic contrast gap that limits quantitative imaging. Approaches that retain low-frequency phase contrast exist but typically require substantially increased experimental complexity, restricting routine use. Dark-field STEM imaging captures this missing low-frequency information through electrons scattered outside the bright-field disk, but discards a large fraction of the scattered signal and is therefore dose-inefficient.
		
		Fused Full-field STEM (FF-STEM) is introduced as a 4D-STEM imaging modality that overcomes this limitation by combining ptychographic phase reconstruction with tilt-corrected dark-field imaging within a single acquisition. Bright-field data are used to estimate probe aberrations and reconstruct a high-resolution phase image, while dark-field data provide complementary low-frequency contrast. The two channels are optimally fused in Fourier space using minimum-variance weighting based on the spectral signal-to-noise ratio, yielding transfer-gap-free images with high contrast and quantitative fidelity. FF-STEM preserves the upsampling and depth-sectioning capabilities of ptychography, adds robust low-frequency contrast characteristic of dark-field imaging, and enables dose-efficient, near–real-time reconstruction.
	\end{abstract}

	\section*{Introduction}
	Transmission electron microscopy (TEM) is a central technique for nanoscale materials characterization, providing imaging with atomic-level resolution. In scanning transmission electron microscopy (STEM), advances in direct electron detector technology \cite{levin2021direct, tate2016high} now enable the acquisition of a two-dimensional convergent beam electron diffraction pattern at every probe position, giving rise to four-dimensional STEM (4D-STEM) \cite{ophus2019four, philipp2022very}. Due to the rich information encoded in the 4D-STEM dataset, a variety of reconstruction strategies have been developed to extract structural and phase information of the specimen. Among these methods, iterative ptychography \cite{maiden2009improved, jiang2018electron} delivers the highest quantitative accuracy and resolution, as it models multiple scattering and probe-sample interactions through successive forward and backward propagations of the electron wave. Combining with multislice \cite{chen2021electron} or tomography \cite{pelz2023solving, romanov2024multi}, these iterative schemes have demonstrated sub-Ångstrom phase retrieval on bulk-like materials and even three-dimensional reconstructions of complex specimens with sub-\AA ngstrom 3D resolution \cite{you2024near}.
	
	Simple, real-time capable approaches such as the center-of-mass (COM) \cite{muller2017measurement, close2015towards, lazic2016phase}, differential phase contrast (DPC) \cite{muller2019comparison} or optimum bright-field (OBF) STEM \cite{ooe2025dose} methods retrieve the projected electrostatic potential using linear approximations to the phase-retrieval problem.\\
	More advanced techniques enable analytical phase retrieval of the specimen transmission function from the recorded interference between overlapping diffraction disks under the weak-phase approximation. Direct ptychography methods correct aberrations with a Fourier filter operation before aggregating all information scattered into the bright field regions into a 2D image \cite{rodenburg1992theory,pennycook2015efficient,ooe2025dose,yang2016simultaneous,ma2025information}.
	
	Tilt-corrected bright-field \cite{nguyen20164d, spoth2017dose, yu2025dose} or parallax imaging \cite{varnavides2023iterative}, enhance phase-contrast transfer at spatial frequencies smaller than the numerical aperture by compensating for defocus-induced image shifts, while additional aberrations lead to significant damping of the transferred contrast.\cite{ma2025using}. 
	These linear bright-field reconstruction methods inherently suppress low spatial frequency components of the image, leading to reduced contrast in slowly varying specimen regions.
	
	
	The electrons scattered into the dark-field, on the other hand, are known to transfer low spatial frequencies well, as they mainly contain incoherent signals \cite{Hartel_Rose_Dinges_1996,nellist2000principles} when a large detector is used. Tilt-corrected dark-field (tcDF) imaging was recently introduced \cite{ma2025information} to fully utilize the detected electrons, also with linear imaging methods. By grouping high-angle scattered electrons into segmented dark-field regions and compensating for defocus-induced image shifts, tcDF reconstructs images that emphasize slowly varying structural features and mass–thickness contrast. This approach is robust for thick specimens or regions affected by dynamical scattering, where bright-field-based methods often fail to provide accurate intensity transfer. However, while tcDF enhances low-frequency contrast, it does not recover phase information accessible through ptychography. Therefore, combining the complementary strengths of direct ptychography and tilt-corrected dark-field imaging provides a promising way toward a unified reconstruction framework that preserves both high- and low-frequency contrast in 4D-STEM.
	
	Building upon the complementary contrast mechanisms of bright-field phase-contrast methods and dark-field imaging, we introduce the Fused Full-field STEM (FF-STEM) reconstruction method which integrates information from both direct ptychography and tcDF imaging, while preserving the capability of upsampling, depth sectioning and near-real-time reconstruction. In this approach, the bright-field channel contributes phase information and high-frequency detail, while the dark-field channel restores the low-frequency signal missing from direct ptychography reconstructions. The fusion is performed in Fourier space, guided by minimum variance weighting functions derived from the spectral signal-to-noise ratio (SSNR) of both signal channels, ensuring a smooth and physically consistent transition between spatial frequency regimes. In addition, our implementation incorporates aberration determination through analytical gradient backpropagation using an image quality metric. For megapixel images, all operations are executed in less than half second on modern parallel processing hardware. As a result, FF-STEM achieves high-contrast imaging that unites the strengths of both reconstruction modes while retaining near-real-time performance.
	\section*{Results} 
	\subsection*{Full-Field STEM recovers high-contrast and fully dose-efficient STEM images with gap-free information transfer}
	\begin{figure*}[htbp!]
		\includegraphics[width=\textwidth]{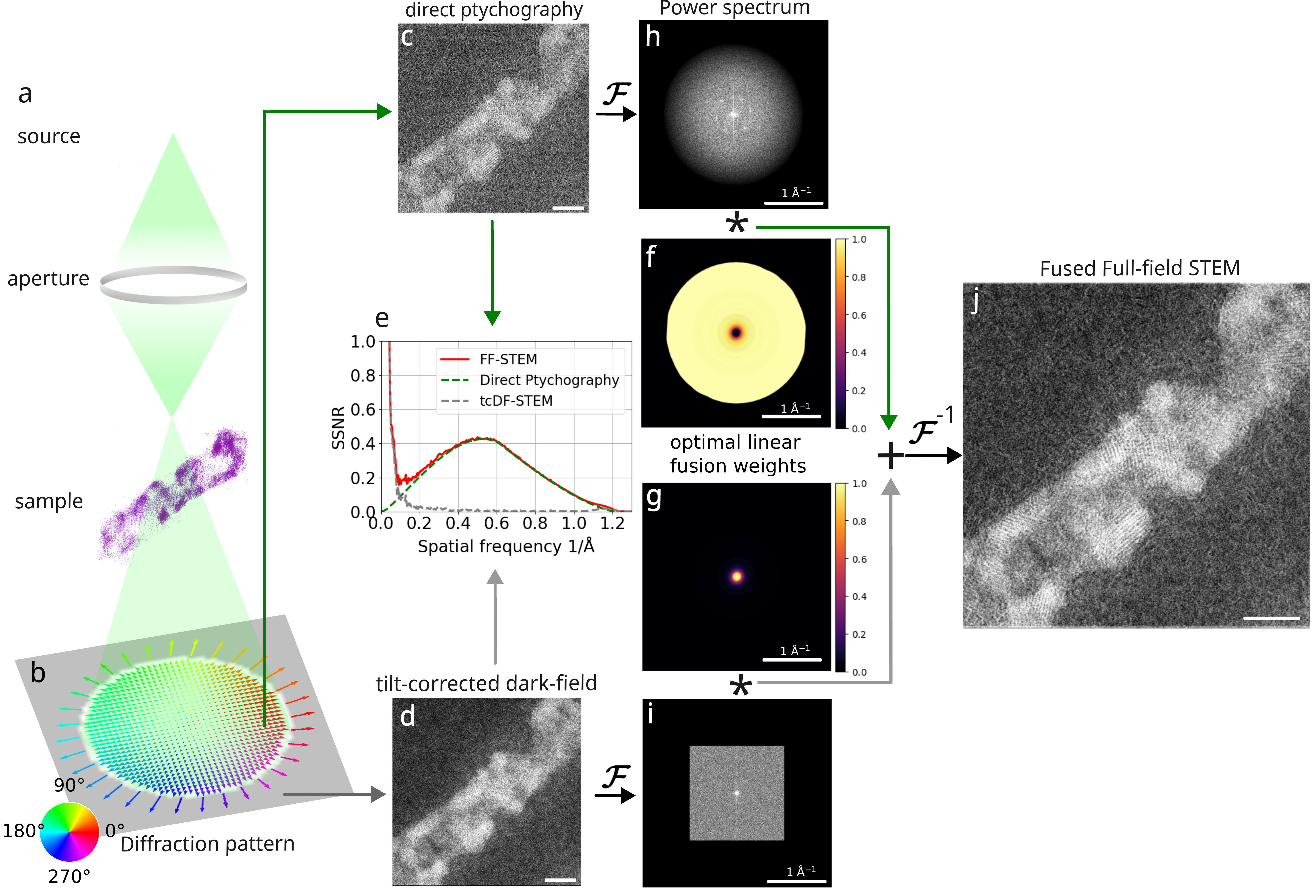}
		\caption{\label{fig:fig1} \textbf{Schematic illustration of the FF-STEM reconstruction workflow} (a) Simplified 4D-STEM acquisition scheme in which a convergent electron probe raster scans through the specimen while a two-dimensional diffraction pattern is recorded at each scan position, the sample is a $\mathrm{Gd_2O_3}$ nanohelix. (b) Example diffraction pattern showing the bright-field (white) and dark-field (gray) regions. The arrows represent the calculated shifts. (c) Aberration-compensated direct ptychography image (d) tcDF image reconstructed by correcting image shifts of the dark field segments indicated by color-coded arrows in (b). Next, the spectral signal-to-noise ratio of both channels is calculated (e) and Wiener-type minimum variance spectral weight for each channel is derived from both SSNRs. (f) and (g) show the spectral weights of ptychography and tcDF respectively. The Fourier spectra of both channels (h,i) are filtered with the optimal weights and are then summed in the Fourier domain and inverse-transformed to yield the final (j) FF-STEM image that utilizes all detected electrons. Scale bars in real space images are 5nm.}
	\end{figure*}
	
	Figure \ref{fig:fig1} illustrates the principle of FF-STEM. A convergent probe interacting with a nano-spiral specimen produces a 2D diffraction pattern containing a bright-field (BF) disk and surrounding dark-field (DF) scattering (Figure \ref{fig:fig1}a). Raster scanning the probe yields a 4D-STEM dataset containing both spatially resolved diffraction and real-space contrast (Figure \ref{fig:fig1}b).
	
	The BF disk contains the coherent interference between the unscattered probe and the first-order diffracted beams \cite{Rose_1976}; it therefore carries the coherent amplitude and phase information used by direct ptychography and other BF-based STEM phase-retrieval methods. However, this channel suffers from a well-known dip at low spatial frequencies because only interference of real-space points lying inside the small probe illumination window is encoded linearly in the diffraction signal \cite{Varnavides_2025}. In contrast, the DF region contains strong low-frequency information arising from incoherent amplitude scattering \cite{Rose_1976,ma2025information}. Simultaneously, azimuthal dark field segments also display image shifts depending on the defocus aberration of the probe. These shifts encode low-frequency and depth information but are usually lost when the DF signal is summed incoherently \cite{ma2025information}.
	
	FF-STEM unifies these two complementary sources of information. Any aberrations present in th BF region are algorithmically determined and corrected using the aberration gradient obtained from an image-contrast metric of the ptychography image (Methods, Equ. \eqref{eq:phase_image_direct_ptycho}), while the DF region is corrected using tilt-corrected dark-field (tcDF) imaging (Methods, Equ. \eqref{eq:tcdf_img}), which removes the defocus-induced real-space shifts associated with each azimuthal DF segment \cite{ma2025information}. After these corrections, both the BF-based ptychographic image and the DF-based tcDF image represent the same underlying object in a common coordinate system.
	
	To combine these reconstructions, we estimate the spectral signal-to-noise ratio (SSNR) of each modality. For ptychography, the SSNR follows a known analytical form that depends only on the probe aperture and aberration surface \cite{Varnavides_2025, Bennemann_Nellist_2025} (Methods, Equ. \eqref{eq:aperture_autocorrelation_eq}). For tcDF, the signal is object-dependent and no analytical SSNR exists \cite{Rose_1976,ma2025information}; we therefore estimate it from two independent half-data tcDF reconstructions using the established half-split approach from cryo-EM \cite{Unser_Trus_Steven_1987,Heel_Schatz_2005} (Methods, Equ. \eqref{eq:ssnr_tcdf}). This provides unbiased estimates of the signal and noise power spectra of the tcDF channel.
	
	The two imaging modalities are then fused in Fourier space using Wiener-type weights derived directly from their SSNRs (Methods, Equ. \eqref{eq:wiener_weights}). At each spatial frequency, the weight automatically selects the channel with the higher spectral reliability: tcDF dominates at low spatial frequencies, where ptychography is intrinsically suppressed, while ptychography dominates at higher frequencies, where tcDF carries little coherent signal. The fusion therefore yields continuous, gap-free contrast transfer across the entire spatial-frequency range.
	
	Let $\widehat{O}(\mathbf{q})$ and $\widehat{O}_{\mathrm{tcDF}}(\mathbf{q})$ denote the Fourier transforms of the ptychography and tcDF reconstructions. The FF-STEM image in Fig. \ref{fig:fig1} j) is obtained by inverse Fourier-transforming the weighted sum of these spectra (Methods, Equ. \eqref{eq:ff_stem_spectrum}). Because the two channels provide statistically independent signal, the SSNR of FF-STEM is simply the sum of the channel-wise SSNRs, providing a direct demonstration of complementary information, shown in Figure \ref{fig:fig1}e.
	
	Applied to the nano-helix dataset in Figure \ref{fig:fig1}, FF-STEM yields a high-contrast image with signal transfer from the lowest spatial frequencies to the edges of twice the numerical aperture limit. The resulting SSNR curve (Figure \ref{fig:fig1}e) shows complete removal of the low-frequency dip characteristic of ptychography and other BF-STEM imaging methods, confirming that FF-STEM restores information that is so far not recovered by bright-field STEM imaging methods.\\

	\subsection*{Contrast transfer and depth sectioning using FF-STEM}
	\subsubsection*{Recovery of low- and high-frequency information in simulation}
	\begin{figure*}[htbp!]
		\includegraphics[width=\textwidth]{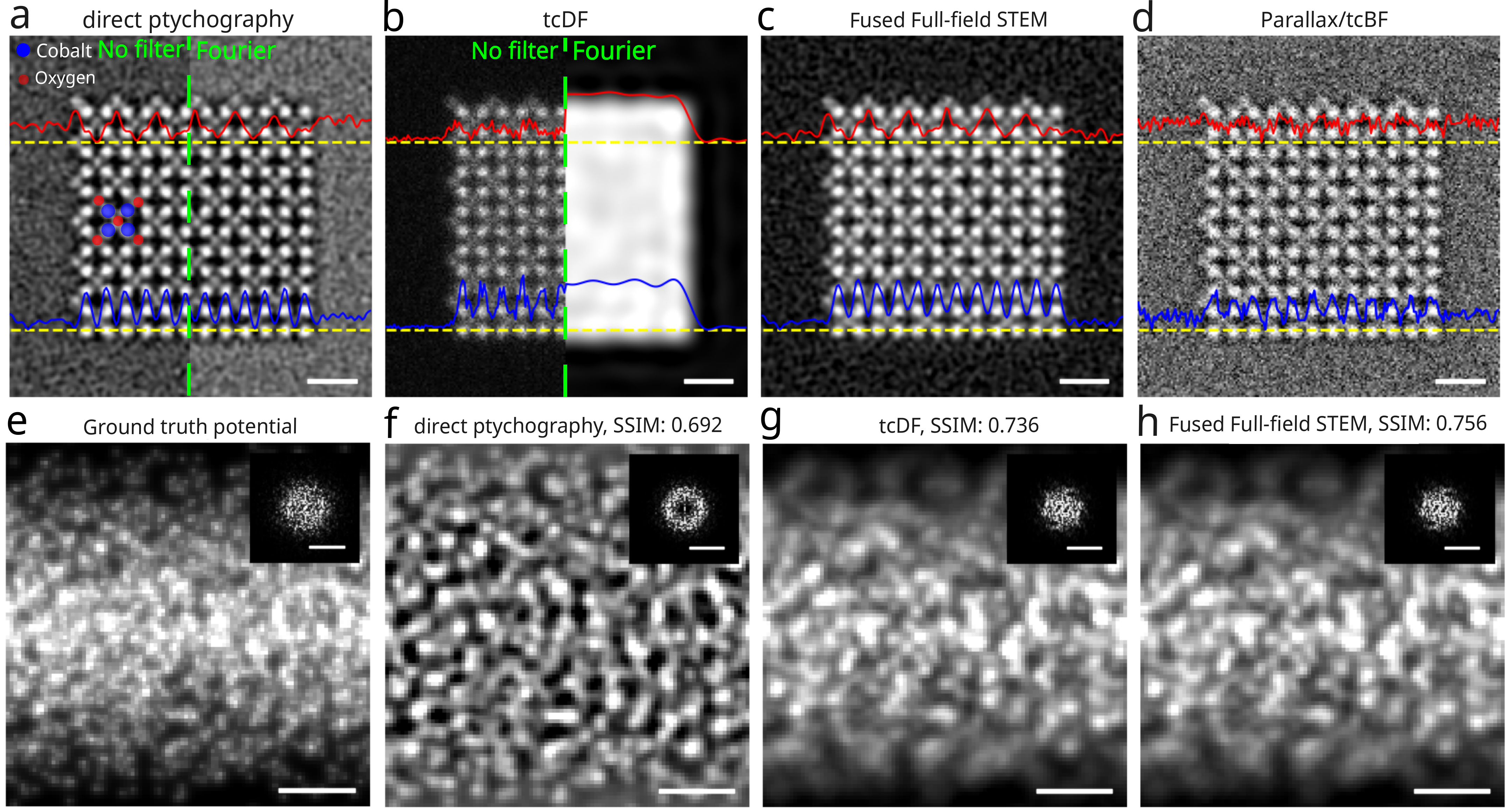}
		\caption{\label{fig:fig2} \textbf{Comparison of direct ptychography, tcDF, FF-STEM, and parallax/tcBF reconstructions}. (a–d) Atomic-resolution reconstructions of a $\mathrm{Co_3O_4}$ nanocrystal. In (a), the projected unit cell is indicated. Panels (a) direct ptychography and (b) tcDF reconstruction. Left half  without and the right half with Wiener-type filters applied. (c) FF-STEM reconstruction. (d) Parallax/tcBF reconstruction. (e–h) Reconstructions of a \SI{7.2}{\nano\meter}-thick amorphous carbon wedge. Real-space scale bars correspond to \SI{5}{\angstrom}, and reciprocal-space scale bars correspond to \SI{1}{\angstrom^{-1}}}.
	\end{figure*}
	
	To evaluate the performance of the FF-STEM method, we simulated a $\mathrm{Co_3O_4}$ nanoparticle with a fluence of \SI{1.0e3}{\elementarycharge\per\angstrom^2}. The aberrations of the 4D-STEM dataset are calculated using our gradient backpropagation approach (Methods, Equ. (8)). With the calculated aberrations, the direct ptychography reconstruction and tcDF are performed independently. For the image fusion we combine the direct ptychography and tcDF using Wiener-type filter weights corresponding to the minimum-variance linear estimator (Methods, Equ. \eqref{eq:wiener_weights}). 
	
	The direct ptychography result (Figure \ref{fig:fig2}a) clearly resolves atomic columns but exhibits attenuated low-frequency contrast, leading to reduced background separation between the nanoparticle and vacuum. The tcDF image (Figure \ref{fig:fig2}b), on the other hand, results in significant contrast difference between nanoparticle and vacuum, but lacks phase contrast and high-frequency detail. The fused image (Figure \ref{fig:fig2}c) combines the strengths of both methods, producing an improved overall contrast with clearer and less noisy high-frequency content, as confirmed by the line profiles across the oxygen (red) and cobalt (blue) rows. The Co rows display sharper and higher-amplitude peaks, while the O rows remain well resolved, indicating faithful transfer of both high- and low- spatial-frequeny components. Compared with the parallax reconstruction in Figure \ref{fig:fig2}d, the FF-STEM method achieves comparable atomic resolution with superior backround suppression, demonstrating its capability for high contrast aberration-free imaging in 4D-STEM.
	
	Figure \ref{fig:fig2}e-h show a quantitative comparison of a simulated amorphous carbon wedge, where the thickness modulation along the vertical spans the full image. Visually, direct ptychography very weakly transfers this large-scale contrast variation, as evidenced by the dip in the inset power spectrum at low spatial frequencies, whereas it is excellently preserved in tcDF-STEM. Quantitatively, we compare the reconstruction quality with the Spatial Similary Index Measure (SSIM). Here, FF-STEM offers the best performance, with a 9.2\% improvement over direct ptychography.
	
	\subsubsection*{Depth sectioning from a single 4D-STEM dataset}

	\begin{figure*}[htbp!]
		\includegraphics[width=\textwidth]{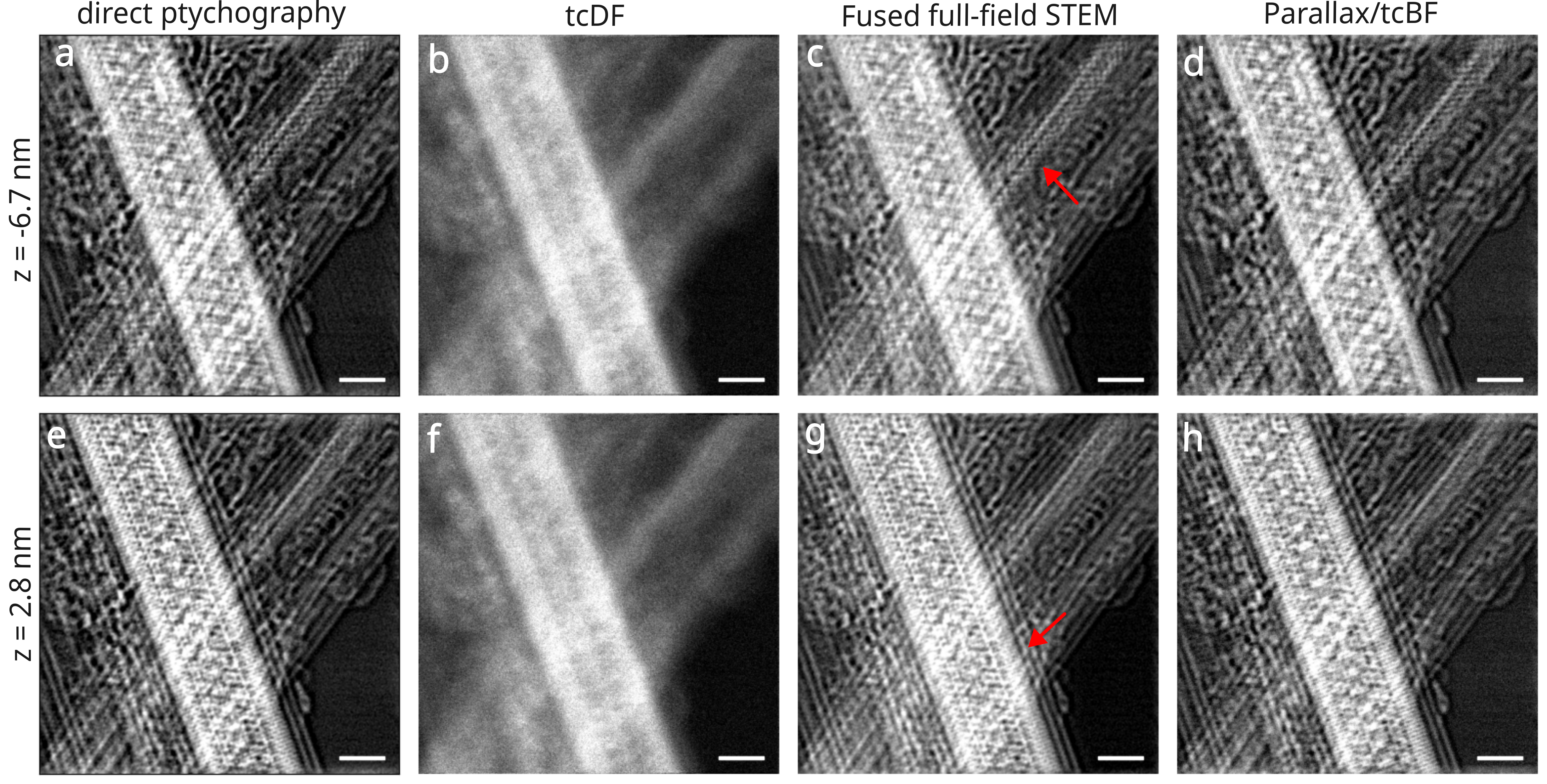}
		\caption{\label{fig:fig3} \textbf{Demonstration of depth-sectioning capability using FF-STEM reconstruction.} Reconstructed images of Ta-Te core-shell carbon nanotubes with different defocus values. Each column shows results from (a, e) direct ptychography, (b, f) tcDF, (c, g) FF-STEM, and (d, h) parallax reconstruction. The top row corresponds to a defocus condition in which the reconstruction focuses on the nanotube bundle at the back of the sample, while the bottom row focuses on the front bundle. The parallax reconstruction provides a comparable focal selectivity but lower overall image clarity. Scale bars: 2nm}
	\end{figure*}
	
	To demonstrate the FF-STEM method's three-dimensional imaging capability on experimental data, we applied it to published dataset of a Ta-Te core-shell carbon-nanotube sample \cite{pelz2021phase}. By systematically varying the defocus value during reconstruction, we obtain depth-resolved images in which different axial planes of the specimen come into focus. In the upper row of Figure \ref{fig:fig3}, the reconstruction is focused on the nanotube bundle located toward the back of the sample, whereas in the lower row, the focal plane shifts to the front bundle. Among the individual methods, direct ptychography shows distinguishable depth separation between the two focal planes. At the same time, the tcDF reconstructions appear blurred and lack high-frequency detail, making the focal shift less evident. This is due to the relatively low electron count in the dark field region, which was only recorded up to twice the semiconvergence angle on the 4Dcamera detector. In contrast, the FF-STEM reconstruction maintains sharp lattice contrast and consistent background noise suppression across depths, indicating robust transfer of information over a broad range of spatial frequencies. The parallax reconstruction shown for comparison produces focal selectivity similar to that shown by direct ptychography. These results confirm that the FF-STEM approach retains the ability to retrieve 3D information and enables near-real-time depth sectioning within a single 4D-STEM acquisition.
	We note that this linear depth sectioning does not model multiple scattering, unlike modern nonlinear multi-slice ptychography (MSP) algorithms \cite{chen2021electron}, but rather provides a near-real-time way to gain understanding of the sample geometry and z-positioning without solving the inverse-multislice problem. This capability could, for example, be used to initialize MSP reconstructions with reasonable hyperparameters, which typically converge in minutes to tens of minutes with current implementations and single-node computational resources.
	
	\subsection*{FF-STEM for dose-efficient high-contrast imaging in the materials sciences}
	
	\begin{figure*}[htbp!]
		\includegraphics[width=\textwidth]{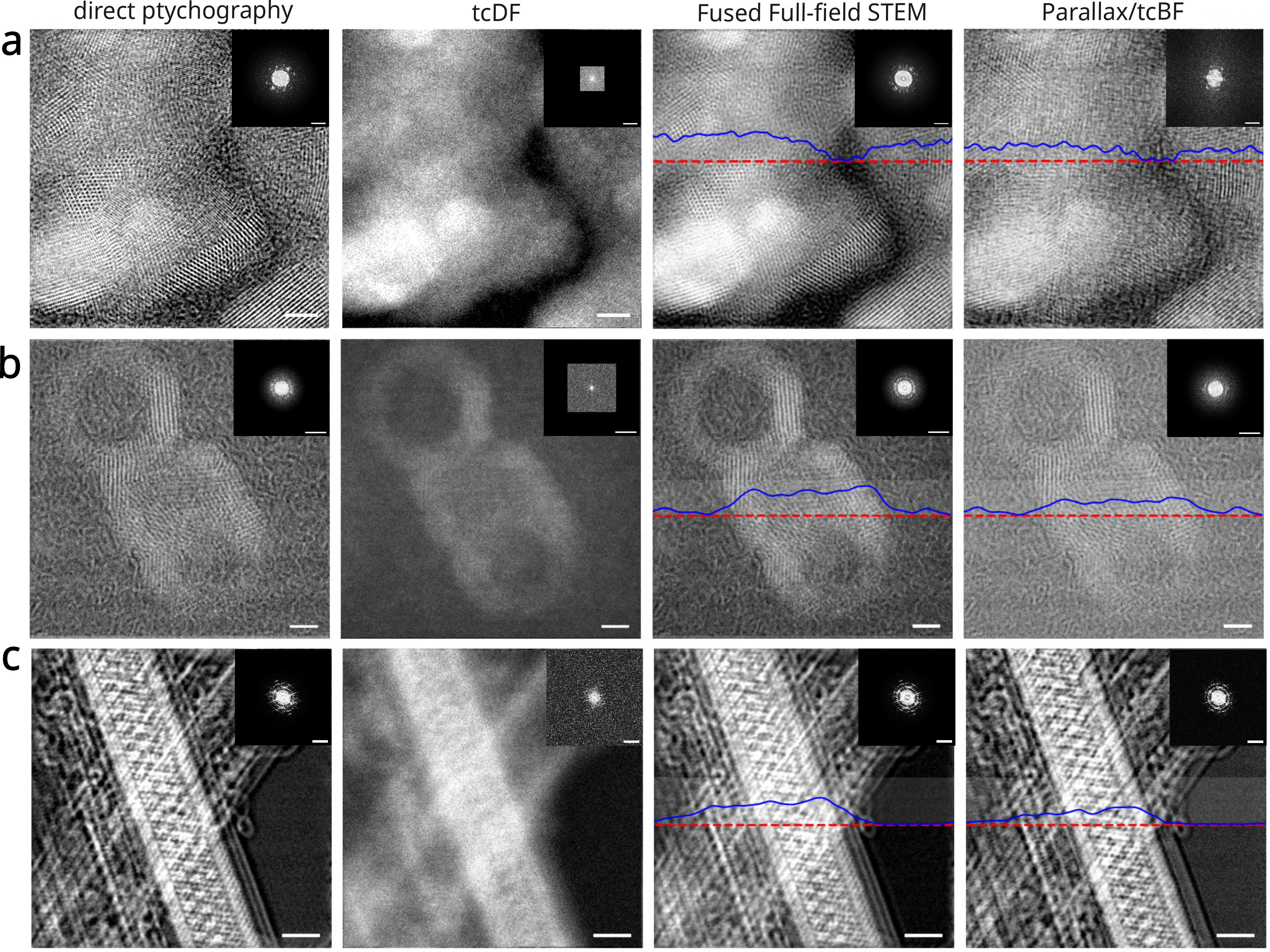}
		\caption{\label{fig:fig4} \textbf{Experimental demonstration of FF-STEM on various 4D-STEM datasets.} Each row represents a distinct experimental dataset obtained from different materials and detectors. The FF-STEM results are compared with parallax reconstruction, showing the contrast improvement of the nanoparticle with line profiles overlaid on the images. (a) Reconstructed images of the diffraction grating. All reconstructions, including parallax, are upsampled by a factor of 4. (b) $\mathrm{Gd_2O_3}$ nanohelices acquired with the Timepix4 detector under low dose conditions, all reconstructions are upsampled by 2. (c) carbon nanotubes with Ta-Te core-shell structure. The red dashed line in the fused FF and parallax images mark the region used to extract the line profiles (blue curve). The scale bars in real space correspond to 2nm. Insets on the top right show the power spectrum of the corresponding images with reciprocal-space scale bars of \SI{0.2}{\angstrom^{-1}} for diffraction grating, and \SI{0.5}{\angstrom^{-1}} for $\mathrm{Gd_2O_3}$, Ta-Te.}
	\end{figure*}
	
	We further applied FF-STEM reconstruction to three datasets spanning different materials systems and detector configurations. The first dataset, as shown in Figure \ref{fig:fig4}a, is a standard diffraction-grating replica generally used for TEM alignment. The dataset is acquired at 200 kV acccelerating voltage, 30 mrad convergence semiangle, 30 pA beam current, and 8 \SI{}{\micro\second} dwell time with a 0.727 \SI{}{\angstrom} scan step. All reconstructions, direct ptychography, tcDF, FF-STEM, and parallax are performed with a four-fold upsampling to the Nyquist limit, with tcDF and parallax upsampled by zero-padding in Fourier space, and direct ptychography upsampled by tiling in Fourier space, explained in Methods (Equ. \eqref{eq:upsample_direct_ptycho}). In the upsampled reconstructions, the FF-STEM image exhibits enhanced lattice continuity and smoother contrast modulation, accompanied by reduced background noise in both real and reciprocal space. The upsampled parallax reconstruction, while capturing the overall periodicity, shows relatively noisy background. The corresponding power spectra show that the FF-STEM image recovers more Bragg peaks and exhibits a lower diffuse background intensity. The line profile taken along the red dashed region in the FF-STEM image further illustrates the improved signal-to-noise ratio achieved by the FF-STEM approach. 
	
	The second dataset in Figure \ref{fig:fig4}b, showing $\mathrm{Gd_2O_3}$ \cite{liu2020colloidal} nanohelices, is recorded at 60 keV electron energy and a 30 mrad semiconvergence angle using a Timepix4 detector at a fluence of $1211 \mathrm{e/\angstrom^2}$. It showcases the method's robustness under low-count and very fast dwell-time conditions of a modern event-based detector using 1 \SI{}{\micro\second} dwell time and 36pA beam current. With an upsampling factor of 2, the reconstructed image clearly resolves atomic features at a Nyquist resolution of 0.86\SI{}{\angstrom}, while the corresponding power spectrum inset reveals distinct Bragg reflections and reduced low-frequency background, confirming that the fusion suppresses the noise from low-count binary datasets \cite{oleary2020phase}. Such event-based datasets recorded with ultrafast dwell times can be beneficial for very low-dose imaging of beam-sensitive materials \cite{yuan2025atomically}, and might enable motion correction for cryo-electron ptychography \cite{huang2023cryo}.
	
	Finally, we applied FF-STEM a different region of Ta-Te core-shell carbon nanotubes \cite{pelz2021phase}, acquired under multiple-scattering conditions with the 4Dcamera at 80 keV and 25mrad semi-convergence angle. The FF-STEM reconstruction enhances both the high-frequency periodicity of the Ta-Te lattice and the low-frequency envelope contrast of the surrounding carbon walls. Across all four experiments, the fused method consistently delivers balanced low- and high-frequency transfer, confirming its generalization beyond simulations and its suitability for high-contrast imaging in diverse experimental settings.
	
	\subsection*{FF-STEM for dose-efficient imaging of thick samples in cryo-electron microscopy}
	\begin{figure*}[htbp!]
		\includegraphics[width=\textwidth]{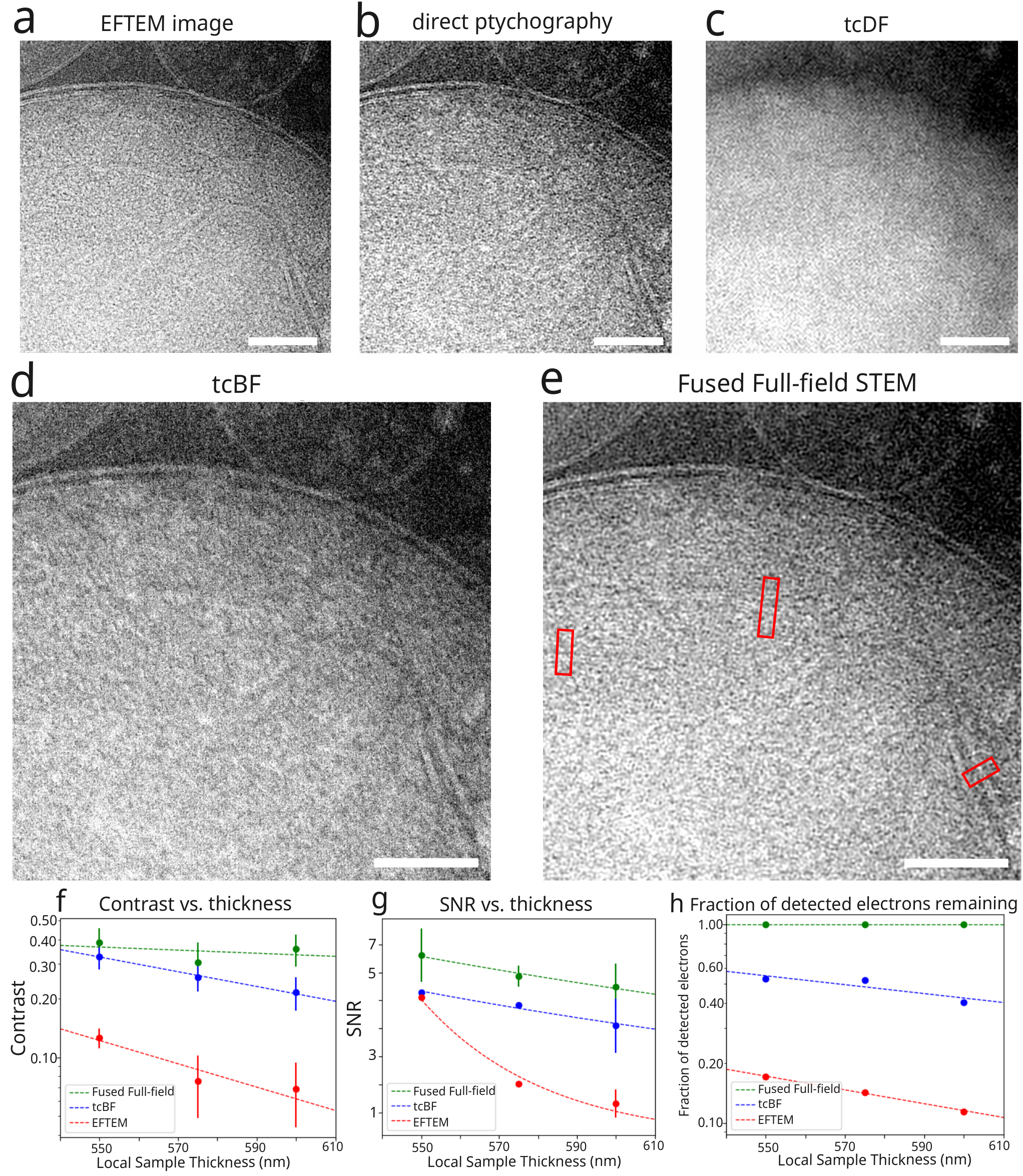}
		\caption{\label{fig:fig5} \textbf{Dose-efficient cryogenic imaging of thick specimens using FF-STEM.} (a) EFTEM image reproduced from \cite{yu2025dose}, acquired at $14 \mathrm{e/\angstrom^2}$. (b) Direct ptychography reconstruction from the 4D-STEM dataset (c) tcDF reconstruction, (d) tcBF image reproduced from \cite{yu2025dose}, (e) FF-STEM reconstruction. Red rectangles indicate regions of interest used for quantitative analysis. (f) Image contrast, (g) signal-to-noise ratio (SNR), and (h) fraction of detected electrons remaining, all extracted from identical regions of interest marked in panel (e). Error bars represent the root-mean-square variation of the measured line profiles. The scale bar is 100nm.}
	\end{figure*}
	
	As a second application, we evaluate FF-STEM for imaging of thick biological materials. Figure \ref{fig:fig5} compares energy-filtered TEM and various 4D-STEM imaging methods on a published dataset, imaging a approximately 600 nm thick region of a mitochondrion at 300kV, collected at a fluence of $14 \mathrm{e/\angstrom^2}$\cite{yu2025dose}. As in the original publication, we compare the fringe contrast and signal-to-noise ratio of the membrane bilayer features, as well as the fraction of detected electrons used in the imaging method. 
	
	As in the original publication \cite{yu2025dose}, tcBF imaging shows improved contrast compared to EFTEM, and the fraction of electrons contributing to the image is dramatically increased. Direct ptychography provides similar, but slightly better contrast than tcBF, which was recently also explained from an SSNR point of view \cite{Varnavides_2025,Bennemann_Nellist_2025,ma2025information}. This contrast improvement is carried over to FF-STEM, which also inherits the thickness and low spatial frequency contrast from the tcDF signal. 
	
	FF-STEM consistently exhibits higher contrast and SNR while retaining a substantially larger fraction of detected electrons than tcBF and EFTEM across the full thickness range, reflecting its ability to make use of electrons scattered into both the bright-field and dark-field regions of the detector.
	
	\begin{figure*}[htbp!]
		\includegraphics[width=\textwidth]{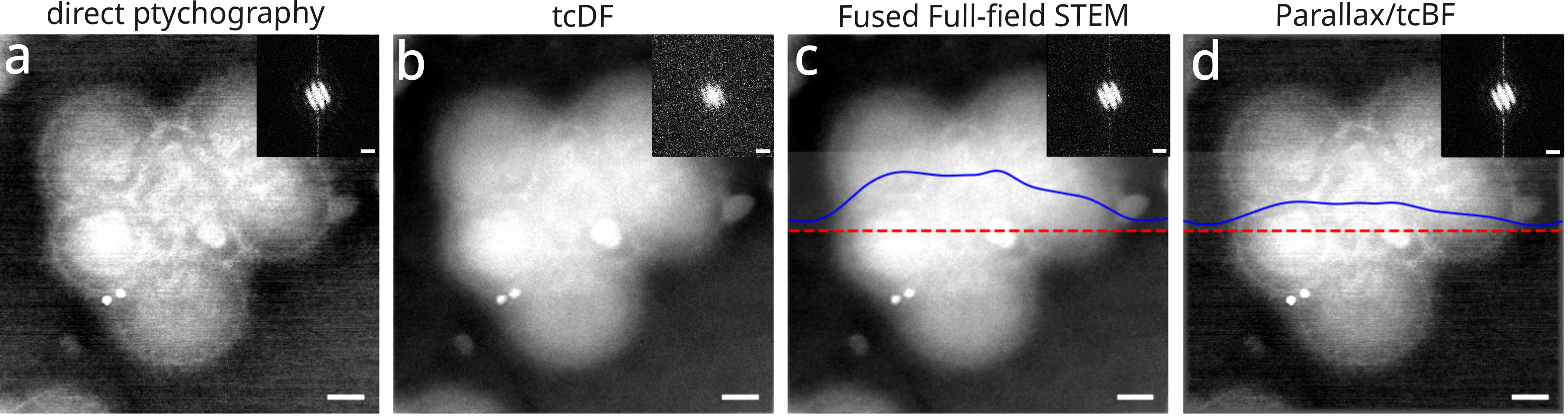}
		\caption{\label{fig:fig51} \textbf{Thickness- and phase-contrast at low fluence and high convergence angles.} Vitrified Virus-like particle cluster imaged under cryogenic conditions with a fluence of $32 \mathrm{e/\angstrom^2}$ and 30 mrad convergence angle. Insets on the top right show the power spectrum of the corresponding images. The scale bar in real space is 40nm. in reciprocal-space is 0.1\SI{}{\angstrom^{-1}}.}
	\end{figure*}
	
	The last dataset shown in Figure \ref{fig:fig51} demonstrates FF-STEM imaging of a virus-like particle (VLP) and is recorded at 200 keV electron energy and a 30 mrad semiconvergence angle using a Dectris ARINA detector with \SI{10}{\micro\second} dwell time at a fluence of $32 \mathrm{e/\angstrom^2}$, highlighting the method's capability to recover mesoscale structure with strong low-frequency components. The reconstructed image preserves the particle morphology and fine structural variations despite the relatively large field of view, demonstrating the method's adaptability across length scales. Large convergence angles allow depth resolution of a few nanometers, but such high convergence angles strongly dampen large-scale features across the whole virus particle in ptychographic reconstructions. These features are preserved in the dark-field signal and subsequently in FF-STEM. They can be used in the future to provide excellent depth information in limited-angle tomography experiments, similar to recent approaches that exploit so-called shadow montages \cite{Seifer_Houben_Elbaum_2025} to extract 3D information from single 4D-STEM datasets acquired for tomography.
	
	\section*{Discussion}
	
	In summary, we have developed the Fused Full-field 4D-STEM method that combines the complementary strengths of direct ptychography and tcDF imaging within a single, analytical framework. The approach simultaneously enhances low- and high-spatial-frequency contrast, enables automated aberration determination and compensation, and supports Fourier-domain upsampling and depth sectioning, all while maintaining near-real-time performance. Through simulations and multiple experimental datasets spanning nanocrystalline oxides, carbon nanotubes, and thick biological samples, FF-STEM yields high-contrast images across diverse materials systems and detector configurations. The method preserves atomic-resolution detail, reveals specimen depth information, and achieves complete reconstruction of multi-gigabyte 4D-STEM data within half second with current parallel processing hardware. 
	
	These results establish FF-STEM as a versatile, computationally efficient tool for high-contrast, highly dose-efficient STEM imaging and open the way for near-real-time feedback and adaptive microscopy in complex electron-scattering experiments. 
	\newpage
	\section*{Methods} 
	
	\subsection*{Upsampled Direct Ptychography}
	The Fourier transform of a 4D-STEM dataset in the projection approximation can be described as the function $G(\mathbf{k}, \mathbf{q})$ \cite{Yang_2016}
	\begin{equation}
		G(\mathbf{k}, \mathbf{q}) = 
		|A(\mathbf{k})|^2 \delta(\mathbf{q})
		+ A(\mathbf{k}) A^*(\mathbf{k} + \mathbf{q}) O^*(-\mathbf{q})
		+ A^*(\mathbf{k}) A(\mathbf{k} - \mathbf{q}) O(+\mathbf{q}).
	\end{equation}
	Here we use the Fourier-space complex aperture function
	\begin{equation}
		A(\mathbf{k}) \;=\; H(\mathbf{k})\,e^{-\,i\,\chi(\mathbf{k})},
		\label{eq:probe_forming_aperture_eq}
	\end{equation}
	where \(H(\mathbf{k})\) is the aperture amplitude (top-hat in the implementation) and \(\chi(\mathbf{k})\) is the aberration phase. 
	Using the weak-phase approximation, it can be shown that 
	\begin{equation}
		G(\mathbf{k}, \mathbf{q}) = |A(\mathbf{k})|^2 \delta(\mathbf{q}) + \Gamma(\mathbf{k}, \mathbf{q}) O(\mathbf{q})
	\end{equation}
	where $\Gamma$ is the so-called aperture-overlap function:
	\begin{equation}
		\Gamma(\mathbf{k}, \mathbf{q}) = A(\mathbf{k}) A(\mathbf{q} - \mathbf{k}) e^{-i \chi(\mathbf{q} - \mathbf{k})} e^{i \chi(\mathbf{k})}
		- A(\mathbf{k}) A(\mathbf{q} + \mathbf{k}) e^{i \chi(\mathbf{q} + \mathbf{k})} e^{-i \chi(\mathbf{k})}
	\end{equation}
	If the aberrations are fixed and known, a bright-field phase-contrast image can then be calculated directly from $G(\mathbf{k}, \mathbf{q})$: 
	\begin{equation}
		O(\mathbf{r}) =  \Im\left[\mathcal{F}^{-1}_{q \rightarrow r}\left[
		\sum_{\mathbf{k}} 
		G(\mathbf{k}, \mathbf{q})  
		\frac{\Gamma^*(\mathbf{k}, \mathbf{q})}{
			|\Gamma(\mathbf{k}, \mathbf{q})|}
		\right]\right],
		\quad 
		\text{for } 
		\mathbf{k} \in \{ |A(\mathbf{k})| \neq 0 \}.
		\label{eq:phase_image_direct_ptycho}
	\end{equation}
	Recently, the possibility of upscaling the reconstructed image beyond the used scan step was introduced by either 
	tiling in Fourier space \cite{varnavides2025relaxing} or interleaving zeros in real-space \cite{yu2025dose}. Here we use the approach by \cite{varnavides2025relaxing} as it reduces the computational complexity of the reconstruction.
	\begin{equation}
		G'(\mathbf{k}, \mathbf{q}) = \mathrm{tile}_{\mathsf{f}} \left[  G(\mathbf{k}, \mathbf{q}) \right]
	\end{equation}
	The final weak-phase image is then computed with upsampled coordinates using summation over the bright-field pixels and correcting aberrations first by multiplying with the aperture overlap function $\Gamma$
	\begin{equation}
		\mathrm{O}(\mathbf{r}) =  \Im\left[\mathcal{F}^{-1}_{q \rightarrow r}\left[
		\sum_{\mathbf{k}} 
		G'(\mathbf{k}, \mathbf{q})  
		\frac{\Gamma^*(\mathbf{k}, \mathbf{q})}{
			|\Gamma(\mathbf{k}, \mathbf{q})|}
		\right]\right],
		\quad 
		\text{for } 
		\mathbf{k} \in \{ |\Gamma(\mathbf{k}, \mathbf{q})| \neq 0 \}.
		\label{eq:upsample_direct_ptycho}
	\end{equation}
	\subsection*{Aberration determination through backpropagation of image-contrast gradients}
	
	We find the aberrations by maximizing an image contrast metric and backpropagating the gradients of this metric through the direct ptychography reconstruction to the aberration coefficients. Namely, we optimize the Total Variation loss function with respect to the 12 cartesian aberration coefficients.
	
	\begin{equation}
		\hat{a}_{1...12} = \arg\min_{{a}_{1...12}} \left( -\,\mathrm{TV}(O(\mathbf{r}) \right)
	\end{equation}
	This approach is known to work also in the visible light regime \cite{loetgering2020zpie}, albeit using an iterative reconstruction algorithm.
	\subsection*{Analytical gradients of the ptychographic aberration parameters}
	In our implementation we use cartesian aberration coefficients \cite{krivanek_1999} to avoid evaluating transcendental functions. With \(u=k_x\lambda\), \(v=k_y\lambda\), we define the aberration function
	\[
	\chi(\mathbf{k}) = \frac{2\pi}{\lambda}\sum_{j=0}^{11} a_j\,\phi_j(u,v),
	\]
	with the basis functions $\phi_j(u,v)$
	\[
	\begin{aligned}
		\phi_0&=\tfrac12(u^2+v^2),\quad
		&\phi_1&=\tfrac12(u^2-v^2),\quad
		&\phi_2&=uv,\\
		\phi_3&=\tfrac13(u^3+uv^2),\quad
		&\phi_4&=\tfrac13(v^3+u^2v),\\
		\phi_5&=\tfrac13(u^3-3uv^2),\quad
		&\phi_6&=\tfrac13(3u^2v-v^3),\\
		\phi_7&=\tfrac14(u^4+v^4+2u^2v^2),\quad
		&\phi_8&=\tfrac14(u^4-v^4),\\
		\phi_9&=\tfrac14(2u^3v+2uv^3),\quad
		&\phi_{10}&=\tfrac14(u^4-6u^2v^2+v^4),\\
		\phi_{11}&=\tfrac14(4u^3v-4uv^3).
	\end{aligned}
	\]
	where we map the coefficients $a_j$ to the known aberration coefficients in the following way: $a_0 = C_{1}$, $a_1 = C_{1,2a}$, $a_2 = C_{1,2b}$, $a_3 = C_{2,1a}$, $a_4 = C_{2,1b}$, $a_5 = C_{2,3a}$, $a_6 = C_{2,3b}$, $a_7 = C_{3}$, $a_8 = C_{3,2a}$, $a_9 = C_{3,2b}$, $a_{10} = C_{3,4a}$, $a_{11} = C_{3,4b}$.
	Therefore
	\begin{equation}
		\frac{\partial\chi(\mathbf{k})}{\partial a_j} \;=\; \frac{2\pi}{\lambda}\,\phi_j(u,v).
	\end{equation} 
	Given a real loss \(L\) with upstream adjoint \(\Delta(\mathbf{q},\mathbf{k})=\partial L/\partial G_{\text{out}}(\mathbf{q},\mathbf{k})\), with $G_{\text{out}}$ being the aberration-corrected G-function , for any real parameter \(p\),
	\begin{equation}
		\frac{\partial L}{\partial p}
		= \sum_{\mathbf{q},\mathbf{k}}
		\Re\!\left\{\overline{\Delta(\mathbf{q},\mathbf{k})}\;\frac{\partial G_{\text{out}}(\mathbf{q},\mathbf{k})}{\partial p}\right\}
		= \sum_{\mathbf{q},\mathbf{k}}
		\Re\!\left\{\overline{\Delta}\;G_{\text{in}}\;\overline{\frac{\partial \Gamma}{\partial p}} \right\},
	\end{equation}
	with $G_{\text{in}}$ the uncorrected G-function. The derivative $\frac{\partial \Gamma}{\partial p}$ is given in the supplementary materials.
	Combining (9) and $\frac{\partial \Gamma}{\partial p}$ (13) yields
	\begin{equation}
		\boxed{\displaystyle
			\frac{\partial L}{\partial a_j}
			= \sum_{\mathbf{q},\mathbf{k}}
			\Re\!\left\{ \overline{\Delta(\mathbf{q},\mathbf{k})}\;
			G_{\text{in}}(\mathbf{q},\mathbf{k})\;
			\overline{\, i \!\left[
				C_-(\partial\chi_0-\partial\chi_-)
				+ C_+(\partial\chi_0-\partial\chi_+)
				\right] } \right\},}
	\end{equation}
	where \(C_-=\overline{A(\mathbf{k})}\,A(\mathbf{k}-\mathbf{q})\) and \(C_+=A(\mathbf{k})\,\overline{A(\mathbf{k}+\mathbf{q})}\), and \(\partial\chi_{\{\cdot\}}=\partial\chi(\mathbf{k}\{\cdot\})/\partial a_j\) from (3).
	
	Here, we use a hard aperture, which simplifies gradient calculation. For a soft aperture, product-rule terms proportional to \(\partial H/\partial a_j\) at \(\mathbf{k}\), \(\mathbf{k}\pm\mathbf{q}\) must be included. In our implementation: complex \(i\) is the \(90^\circ\) rotation \((x,y)\mapsto(-y,x)\) on the \(\mathbb{R}^2\) view of \(\mathbb{C}\). Accumulation uses only the real parts \(\Re\{\overline{\Delta}\,\cdot\}\).
	We implement this gradient calculation with efficient customized CUDA kernels. The main implementation challenge is the summation over both $\mathbf{q}$ and $\mathbf{k}$, which is inefficient when implemented with a trivial atomic addition on massively parallel architectures. Here we opt for a tile-wise sum, where each core performs summation over it's $(\mathbf{q},\mathbf{k})$ components before the global summation.
	
	\subsection{Analytical SSNR of Direct Ptychography}
	
	Under the weak phase object approximation (WPOA), the ptychographic
	contrast transfer function is specimen-independent and depends only on
	the probe wavefunction
	\begin{equation}
		\psi(\mathbf{k}) = A(\mathbf{k})\, e^{-i\chi(\mathbf{k})},
	\end{equation}
	where \(A(\mathbf{k})\) is the normalized top-hat aperture and
	\(\chi(\mathbf{k})\) is the aberration phase parameterized by Seidel
	coefficients.
	
	Following Refs.~\cite{Varnavides_2025, Bennemann_Nellist_2025}, the
	dose-normalized spectral signal-to-noise ratio (SSNR) is
	\begin{equation}
		\mathrm{SSNR}_{\mathrm{ptycho}}(\mathbf{q})
		=
		\frac{
			\displaystyle
			\sum_{\mathbf{k}}
			\left|
			\psi^{*}(\mathbf{k})\psi(\mathbf{q}-\mathbf{k})
			-
			\psi(\mathbf{k})\psi^{*}(\mathbf{q}+\mathbf{k})
			\right|
		}{
			2\,\sqrt{N_{\mathrm{DO+TO}}(\mathbf{q})}
		},
		\label{eq:aperture_autocorrelation_eq}
	\end{equation}
	where \(N_{\mathrm{DO+TO}}(\mathbf{q})\) is the number of pixels in the
	double-overlap and triple-overlap regions of the aperture
	autocorrelation at spatial frequency \(\mathbf{q}\).

	\subsection*{Tilt-corrected dark-field STEM}
	Tilt-corrected dark-field STEM was introduced by Ma et al. \cite{ma2025information}. Starting from Eq.~(8) of \cite{nellist2000principles}, the usual ADF incoherent model $I(\mathbf r)=|P|^2*O(\mathbf r)$ follows when the detector integral becomes independent of the incident angles over the aperture-overlap domain, that is, for a geometrically large annular detector.
	A small dark-field sector violates this condition, so the corresponding
	$I_m(\mathbf r)$ retains coherent dependencies on
	$(\boldsymbol\theta,\boldsymbol\theta')$ and does not reduce to a simple
	intensity convolution.
	In the 4D-STEM contrast decomposition of \cite{ma2025information},
	however, only the incoherent amplitude term contributes in the
	dark-field region.
	A practical tcDF image is obtained by azimuthally shifting each
	dark-field sector by
	$\Delta\boldsymbol\rho_m=(\Delta f+\Delta z)(2\alpha/3)\mathbf u_{\varphi_m}$
	and then summing over all sectors, which preferentially aggregates this
	incoherent channel while restoring the bright-field-like parallax. 
	The resulting tcDF image therefore shows partially coherent amplitude contrast and depth-sectioning behaviour.
	It approaches the ADF-like $|P|^2$ convolution only when the summed sector
	coverage effectively behaves like a large annulus, or when the scattering is sufficiently localized to render the detector integral effectively independent of the incident angles. The following steps describe the practical procedure to compute $I_{\mathrm{tcDF}}$ from a 4D-STEM dataset.
	\begin{algorithm}[H]
		\caption*{Computation of the tcDF-STEM image from 4D-STEM data}
		\begin{algorithmic}[1]
			\Require 4D dataset $I(\mathbf r,\boldsymbol\Theta)$
			\Require detector angle maps $(\Theta_x,\Theta_y)$
			\Require convergence semi-angle $\alpha$
			\Require DF annulus $[\beta_{\min},\beta_{\max}]$
			\Require defocus $\Delta f$
			\Require depth offset $\Delta z$
			\Require scan pixel size $s_{\mathrm{scan}}$
			
			\Statex 
			
			\State \textbf{1. Partition the detector.}
			Divide the dark-field region into $M$ narrow angular sectors of width 
			$\Delta\varphi=2\pi/M$:
			\begin{equation}
				D_m(\boldsymbol\Theta)
				= W_{\rm DF}(|\boldsymbol\Theta|)\,
				\Pi_{\Delta\varphi}\!\big(\arg\boldsymbol\Theta-\varphi_m\big),
			\end{equation}
			with $W_{\rm DF}=1$ for $\beta_{\min}\!\le|\boldsymbol\Theta|\!\le\beta_{\max}$ and $0$ otherwise.
			
			\State \textbf{2. Integrate each sector.}
			For every sector $m$, integrate the 4D intensity over the detector pixels belonging to
			that sector to obtain a real-space image
			\begin{equation}
				I_m(\mathbf r)=\int D_m(\boldsymbol\Theta)\,I(\mathbf r,\boldsymbol\Theta)\,d\boldsymbol\Theta.
			\end{equation}
			
			\State \textbf{3. Compute the tcDF parallax shift.}
			For each sector $m$ with azimuth $\varphi_m$, calculate the shift vector
			\begin{equation}
				\Delta\boldsymbol\rho_m = 
				\big(\Delta f+\Delta z\big)\,\boldsymbol\theta_{\mathrm{ref}}(\varphi_m),
				\qquad
				\boldsymbol\theta_{\mathrm{ref}}(\varphi_m)=\frac{2\alpha}{3}\,\mathbf u_{\varphi_m},
			\end{equation}
			and convert it to scan-pixel units,
			$\Delta\mathbf p_m = \Delta\boldsymbol\rho_m / s_{\mathrm{scan}}$.
			
			\State \textbf{4. Apply the real-space shift.}
			Shift each sector image $I_m(\mathbf r)$ by $\Delta\mathbf p_m$ in scan coordinates.
			A Fourier-domain shift is recommended to preserve phase fidelity:
			\begin{equation}
				I_m^{\rm shifted}(\mathbf r)
				= \mathcal F^{-1}\!\left\{
				\mathcal F\!\left[I_m(\mathbf r)\right]
				e^{-i2\pi\mathbf Q\cdot\Delta\boldsymbol\rho_m}
				\right\}.
			\end{equation}
			
			\State \textbf{5. Sum the shifted sectors.}
			Sum all shifted sector images to form the tcDF image:
			\begin{equation}
				O_{\mathrm{tcDF}}(\mathbf r)=\sum_{m=1}^M I_m^{\rm shifted}(\mathbf r).
				\label{eq:tcdf_img}
			\end{equation}
			
			\Statex
			\State \textbf{Output:} The tcDF image $O_{\mathrm{tcDF}}(\mathbf r)$, which 
			exhibits amplitude-dominated (incoherent) contrast and enables 
			defocus- or height-dependent depth sectioning through the choice of 
			$(\Delta f+\Delta z)$.
		\end{algorithmic}
	\end{algorithm}
	
	\subsection{Half-Data SSNR Estimation for tcDF-STEM}
	
	The tcDF-STEM signal depends on object-dependent elastic scattering
	amplitudes, and no closed-form SSNR exists; we therefore estimate
	\(\mathrm{SSNR}_{\mathrm{tcDF}}\) using two statistically independent
	half-data reconstructions.
	
	The dark-field detector is divided azimuthally.  We create two
	independent reconstructions by assigning alternating DF segments to two
	groups, reconstructing two tcDF images using the identical algorithm.
	Let
	\begin{equation}
		F_A(\mathbf{q}), \qquad F_B(\mathbf{q})
	\end{equation}
	denote the orthonormal Fourier transforms of the two half-data images.
	
	We define half-sum and half-difference spectra
	\begin{equation}
		S(\mathbf{q}) = \tfrac{1}{2}\bigl(F_A(\mathbf{q}) + F_B(\mathbf{q})\bigr),
		\qquad
		N(\mathbf{q}) = \tfrac{1}{2}\bigl(F_A(\mathbf{q}) - F_B(\mathbf{q})\bigr).
	\end{equation}
	
	The radially averaged noise and total power spectra are
	\begin{equation}
		P_{\mathrm{noise}}(q)
		=
		\left\langle |N(\mathbf{q})|^2 \right\rangle_{|\mathbf{q}| = q},
	\end{equation}
	\begin{equation}
		P_{\mathrm{total}}(q)
		=
		\left\langle |S(\mathbf{q})|^2 \right\rangle_{|\mathbf{q}| = q}.
	\end{equation}
	
	The signal power is obtained by noise subtraction,
	\begin{equation}
		P_{\mathrm{signal}}(q)
		=
		P_{\mathrm{total}}(q) - P_{\mathrm{noise}}(q),
		\qquad
		P_{\mathrm{signal}}(q) \ge 0.
	\end{equation}
	
	The tcDF spectral signal-to-noise ratio is
	\begin{equation}
		\mathrm{SSNR}_{\mathrm{tcDF}}(q)
		=
		\frac{P_{\mathrm{signal}}(q)}{P_{\mathrm{noise}}(q)}.
		\label{eq:ssnr_tcdf}
	\end{equation}
	
	\subsection*{Optimal linear fusion: FF-STEM}
	
	We consider the fusion of two independent reconstructions
	of the same underlying object in Fourier space. At each spatial frequency
	$\mathbf{q}$ we assume a linear imaging model
	\begin{equation}
		F_i(\mathbf{q}) = X(\mathbf{q}) + n_i(\mathbf{q}), \qquad i \in \{1,2\},
	\end{equation}
	where $X(\mathbf{q})$ is the (unknown) true Fourier coefficient, and $n_i(\mathbf{q})$ are
	zero-mean noise terms with
	\begin{equation}
		\mathbb{E}[n_i(\mathbf{q})] = 0, 
		\qquad 
		\mathbb{E}\bigl[|n_i(\mathbf{q})|^2\bigr] = \sigma_i^2(\mathbf{q}),
	\end{equation}
	and $n_1(\mathbf{q})$, $n_2(\mathbf{q})$ are uncorrelated for each $\mathbf{q}$.
	We seek a linear fused estimator of the form
	\begin{equation}
		\hat X(\mathbf{q}) = w_1(\mathbf{q})\,F_1(\mathbf{q}) + w_2(\mathbf{q})\,F_2(\mathbf{q}),
	\end{equation}
	and require unbiasedness with respect to $X(\mathbf{q})$, i.e.
	\begin{equation}
		\mathbb{E}[\hat X(\mathbf{q})] = X(\mathbf{q})
		\quad\Rightarrow\quad
		w_1(\mathbf{q}) + w_2(\mathbf{q}) = 1.
	\end{equation}
	
	\subsubsection*{Gaussian noise model}
	Under the above assumptions the estimation error is
	\begin{equation}
		\hat X(\mathbf{q}) - X(\mathbf{q}) = w_1(\mathbf{q})\,n_1(\mathbf{q}) + w_2(\mathbf{q})\,n_2(\mathbf{q}),
	\end{equation}
	and its variance is
	\begin{equation}
		\mathrm{Var}\bigl[\hat X(\mathbf{q})-X(\mathbf{q})\bigr]
		= |w_1(\mathbf{q})|^2 \sigma_1^2(\mathbf{q}) + |w_2(\mathbf{q})|^2 \sigma_2^2(\mathbf{q}),
	\end{equation}
	because the cross-term vanishes for uncorrelated noise.
	Using $w_2(\mathbf{q}) = 1 - w_1(\mathbf{q})$, we can write
	\begin{equation}
		V\bigl(w_1(\mathbf{q})\bigr)
		= |w_1(\mathbf{q})|^2 \sigma_1^2(\mathbf{q}) + |1-w_1(\mathbf{q})|^2 \sigma_2^2(\mathbf{q}).
	\end{equation}
	Minimizing $V$ with respect to $w_1(\mathbf{q})$ yields
	\begin{equation}
		\frac{\partial V}{\partial w_1}
		= 2 w_1 \sigma_1^2 - 2 (1 - w_1)\sigma_2^2 = 0,
	\end{equation}
	and hence
	\begin{equation}
		w_1(\mathbf{q})\,\sigma_1^2(\mathbf{q}) 
		= \bigl(1 - w_1(\mathbf{q})\bigr)\sigma_2^2(\mathbf{q}).
	\end{equation}
	Solving for $w_1(\mathbf{q})$ and $w_2(\mathbf{q})$ gives the familiar inverse-variance weights
	\begin{equation}
		w_1(\mathbf{q}) 
		= \frac{\sigma_2^{-2}(\mathbf{q})}{\sigma_1^{-2}(\mathbf{q}) + \sigma_2^{-2}(\mathbf{q})},
		\qquad
		w_2(\mathbf{q}) 
		= \frac{\sigma_1^{-2}(\mathbf{q})}{\sigma_1^{-2}(\mathbf{q}) + \sigma_2^{-2}(\mathbf{q})}.
	\end{equation}
	These are the unique unbiased weights that minimize the mean-squared error
	of $\hat X(\mathbf{q})$ (Gauss--Markov theorem).
	
	\subsubsection*{Formulation in terms of SSNR}
	Let $S_X(\mathbf{q}) = \mathbb{E}\bigl[|X(\mathbf{q})|^2\bigr]$ denote the signal power at
	frequency $\mathbf{q}$. The spectral signal-to-noise ratio (SSNR) for reconstruction
	$i$ at frequency $\mathbf{q}$ is defined as
	\begin{equation}
		\mathrm{SSNR}_i(\mathbf{q}) 
		= \frac{S_X(\mathbf{q})}{\sigma_i^2(\mathbf{q})}.
	\end{equation}
	Using this definition we have
	\begin{equation}
		\frac{1}{\sigma_i^2(\mathbf{q})} = \frac{\mathrm{SSNR}_i(\mathbf{q})}{S_X(\mathbf{q})}.
	\end{equation}
	Substituting into the inverse-variance weights above, the factor $S_X(\mathbf{q})$
	cancels, and we obtain
	\begin{equation}
		\label{equ:weights}
		w_1(\mathbf{q})
		= \frac{\mathrm{SSNR}_1(\mathbf{q})}{\mathrm{SSNR}_1(\mathbf{q}) + \mathrm{SSNR}_2(\mathbf{q})},
		\qquad
		w_2(\mathbf{q})
		= \frac{\mathrm{SSNR}_2(\mathbf{q})}{\mathrm{SSNR}_1(\mathbf{q}) + \mathrm{SSNR}_2(\mathbf{q})}.
	\end{equation}
	Thus, at each spatial frequency the optimal linear fusion weights are
	directly proportional to the SSNR of each reconstruction.
	
	With these weights, the noise variance of the fused estimator is
	\begin{equation}
		\sigma_{\mathrm{fused}}^2(\mathbf{q})
		= |w_1(\mathbf{q})|^2 \sigma_1^2(\mathbf{q}) + |w_2(\mathbf{q})|^2 \sigma_2^2(\mathbf{q}),
	\end{equation}
	and a straightforward algebraic simplification gives
	\begin{equation}
		\frac{1}{\sigma_{\mathrm{fused}}^2(\mathbf{q})} 
		= \frac{1}{\sigma_1^2(\mathbf{q})} + \frac{1}{\sigma_2^2(\mathbf{q})}.
	\end{equation}
	Multiplying by $S_X(\mathbf{q})$ shows that the SSNRs add:
	\begin{equation}
		\mathrm{SSNR}_{\mathrm{fused}}(\mathbf{q})
		= \frac{S_X(\mathbf{q})}{\sigma_{\mathrm{fused}}^2(\mathbf{q})}
		= \frac{S_X(\mathbf{q})}{\sigma_1^2(\mathbf{q})} + \frac{S_X(\mathbf{q})}{\sigma_2^2(\mathbf{q})}
		= \mathrm{SSNR}_1(\mathbf{q}) + \mathrm{SSNR}_2(\mathbf{q}).
	\end{equation}
	In other words, with SSNR-weighted fusion, the information content at each
	frequency simply adds across independent reconstructions.
	
	\subsubsection*{Extension to Poisson noise}
	Many electron microscopes are nowadays equipped with direct electron detectors, where the dominant noise source is Poisson-distributed counting noise. Consider, for simplicity, a single scalar degree of freedom (e.g.\ one pixel or one Fourier coefficient)
	measured twice with different modalities and electron doses $g_1$ and $g_2$. The measured
	counts $y_i$ obey
	\begin{equation}
		y_i \sim \mathrm{Poisson}(\lambda_i),
		\qquad
		\lambda_i = g_i\,s,
	\end{equation}
	where $s$ is the underlying (dose-normalized) signal. An unbiased estimator
	of $s$ from measurement $i$ is given by
	\begin{equation}
		X_i = \frac{y_i}{g_i},
		\qquad
		\mathbb{E}[X_i] = s,
	\end{equation}
	with variance
	\begin{equation}
		\mathrm{Var}(X_i)
		= \frac{\mathrm{Var}(y_i)}{g_i^2}
		= \frac{\lambda_i}{g_i^2}
		= \frac{s}{g_i}.
	\end{equation}
	We again consider an unbiased linear fusion
	\begin{equation}
		\hat s = w_1 X_1 + w_2 X_2,
		\qquad
		w_1 + w_2 = 1,
	\end{equation}
	with variance
	\begin{equation}
		\mathrm{Var}(\hat s)
		= w_1^2 \frac{s}{g_1} + w_2^2 \frac{s}{g_2}.
	\end{equation}
	Minimizing this variance under the constraint $w_1 + w_2 = 1$ yields
	\begin{equation}
		w_1 = \frac{g_1}{g_1 + g_2},
		\qquad
		w_2 = \frac{g_2}{g_1 + g_2},
	\end{equation}
	i.e., \ the optimal unbiased estimator is
	\begin{equation}
		\hat s
		= \frac{g_1 X_1 + g_2 X_2}{g_1 + g_2}
		= \frac{y_1 + y_2}{g_1 + g_2},
	\end{equation}
	which coincides with the maximum-likelihood estimator for Poisson data.
	
	More generally, any linear transformation of Poisson counts (e.g.\ formation
	of complex Fourier coefficients or reconstructed images) yields, by the
	central limit theorem, approximately Gaussian noise with a variance that can be computed from the underlying Poisson statistics. Once the variance
	$\sigma_i^2(\mathbf{q})$ of each reconstruction at each frequency $\mathbf{q}$ is known or
	estimated, the above Gaussian derivation applies unchanged. The optimal
	unbiased linear fusion in the presence of Poisson noise is therefore still
	achieved by inverse-variance weighting,
	\begin{equation}
		w_i(\mathbf{q}) \propto \frac{1}{\sigma_i^2(\mathbf{q})},
	\end{equation}
	or, equivalently, by SSNR weighting,
	\begin{equation}
		\label{eq:ssnr_ratio_weights}
		w_i(\mathbf{q}) 
		= \frac{\mathrm{SSNR}_i(\mathbf{q})}{\sum_j \mathrm{SSNR}_j(\mathbf{q})}.
	\end{equation}
	In this sense, the SSNR-based weighting provides the minimum-variance
	unbiased linear fusion of multiple independent reconstructions, both under
	Gaussian additive noise, and when the noise originates from Poisson counting
	statistics.

	\subsection{Wiener-Type Fusion Weights}
	
	For two statistically independent imaging channels with SSNRs
	\(\mathrm{SSNR}_1(\mathbf{q})\) and \(\mathrm{SSNR}_2(\mathbf{q})\),
	the minimum-variance linear estimator assigns weights given in Equ. \ref{equ:weights}. The FF-STEM weights are therefore
	\begin{equation}
		w_{\mathrm{ptycho}}(\mathbf{q})
		=
		\frac{
			\mathrm{SSNR}_{\mathrm{ptycho}}(\mathbf{q})
		}{
			\mathrm{SSNR}_{\mathrm{ptycho}}(\mathbf{q})
			+
			\mathrm{SSNR}_{\mathrm{tcDF}}(\mathbf{q})
		},
		\qquad
		w_{\mathrm{tcDF}}(\mathbf{q})
		=
		1 - w_{\mathrm{ptycho}}(\mathbf{q}).
		\label{eq:wiener_weights}
	\end{equation}
	\subsection{FF-STEM Reconstruction}
	
	Let \(\widehat{O}_{\mathrm{ptycho}}(\mathbf{q})\) and
	\(\widehat{O}_{\mathrm{tcDF}}(\mathbf{q})\) denote the Fourier
	transforms of the ptychographic and tcDF reconstructions.  The fused
	spectrum is
	\begin{equation}
		\widehat{I}_{\mathrm{FF-STEM}}(\mathbf{q})
		=
		w_{\mathrm{ptycho}}(\mathbf{q})\,\widehat{O}(\mathbf{q})
		+
		w_{\mathrm{tcDF}}(\mathbf{q})\,\widehat{O}_{\mathrm{tcDF}}(\mathbf{q}),
		\label{eq:ff_stem_spectrum}
	\end{equation}
	and the FF-STEM reconstruction is obtained by the inverse Fourier
	transform
	\begin{equation}
		\label{eq:ff_stem}
		I_{\mathrm{FF-STEM}}(\mathbf{r})
		=
		\mathcal{F}^{-1}
		\!\left[
		\widehat{I}_{\mathrm{FF}}(\mathbf{q})
		\right](\mathbf{r}).
	\end{equation}
	
	Because the two channels provide statistically independent signal, the
	total spectral SNR is additive:
	\begin{equation}
		\mathrm{SSNR}_{\mathrm{FF-STEM}}(q)
		=
		\mathrm{SSNR}_{\mathrm{ptycho}}(q)
		+
		\mathrm{SSNR}_{\mathrm{tcDF}}(q).
	\end{equation}
	\subsection{Independence of Ptychographic and tcDF Channels}
	
	The SSNR-based weights in Eq.~\eqref{eq:ssnr_ratio_weights} rely on the
	assumption that the ptychographic and tcDF reconstructions provide
	statistically independent measurements of the same underlying object
	Fourier coefficients.  This is well justified for FF-STEM for two
	reasons.
	
	First, the two channels are formed from disjoint subsets of the
	diffraction data.  The ptychographic reconstruction uses only the
	bright-field disc, whereas the tcDF reconstruction is computed from
	azimuthal segments in the dark-field ring.  At the detector level, the
	dominant noise source is shot noise, which is uncorrelated between
	different pixels.  Since the bright-field and dark-field regions are
	non-overlapping, the corresponding noise realizations entering the
	ptychographic and tcDF pipelines are independent.
	
	Second, the reconstruction algorithms are linear with
	respect to the measured intensities, although the contrast mechanims may be nonlinear.
	Ptychography reconstructs the object by solving a linearized inverse
	problem, whereas tcDF reconstructs tilt-corrected dark-field images from angle-resolved intensities using linear operations. Both channels encode the underlying scattering potential, while their noise contributions originate from independent detector pixels and propagate linearly through the reconstruction operators.
	We therefore model the ptychographic and tcDF Fourier coefficients as
	\begin{equation}
		\widehat{O}_{\mathrm{ptycho}}(\mathbf{q}) = O(\mathbf{q}) + n_{\mathrm{ptycho}}(\mathbf{q}),
		\qquad
		\widehat{O}_{\mathrm{tcDF}}(\mathbf{q}) = O(\mathbf{q}) + n_{\mathrm{tcDF}}(\mathbf{q}),
	\end{equation}
	with
	\begin{equation}
		\mathbb{E}\,n_{\mathrm{ptycho}} = \mathbb{E}\,n_{\mathrm{tcDF}} = 0,
		\qquad
		\mathrm{Cov}\!\left[n_{\mathrm{ptycho}}(\mathbf{q}), n_{\mathrm{tcDF}}(\mathbf{q}')\right] = 0
		\quad \forall\,\mathbf{q},\mathbf{q}'.
	\end{equation}
	Under this model, the SSNR-based fusion in Eq.~\eqref{eq:ff_stem}
	is the minimum-variance linear estimator for $O(\mathbf{q})$.
	\subsection*{EF-TEM experimental parameters}
	The data from the published EFTEM dataset \cite{yu2025dose} were recorded at 300kV with a Falcon 4i detector with a Selectrix X energy filter.
	\subsection*{4D-STEM experimental parameters}
	The experiments with the TimePix4 detector were performed on a TFS Spectra (S)TEM microscope equipped with X-CFEG and a probe corrector, operated at 60kV. The 4D-STEM datasets were recorded at a beam current of 36 pA and a dwell time of 1 microsecond.
	
	The experiments using the ARINA detector were performed on a probe-corrected TFS Spectra 30-200 microscope equipped with XCFEG. 
	
	The experiments using the 4DCamera were performed on the TEAM 0.5 microscope at the National Center for Electron Microscopy, Berkeley, a double-corrected TFS Titan microscope. 
	All experimental parameters are listed in Table 1.
	\begin{table}[t]
		\centering
		\caption{Experimental parameters corresponding to Figure.~1–6.}
		\label{tab:experimental_parameters_transposed}
		\setlength{\tabcolsep}{5pt}
		\renewcommand{\arraystretch}{1.2}
		\begin{tabular}{lcccccc}
			\toprule
			Figure
			& Sample
			& Energy (keV)
			& $\alpha$$$ (mrad)
			& Fluence (e/\AA$^2$)
			& Scan step (\AA)
			& Det. pixel (\AA$^{-1}$) \\
			\midrule
			Fig.~1
			& Gd$_2$O$_3$
			& 60
			& 30
			& $1.2\times10^{3}$
			& 0.61
			& 0.034 \\
			
			Fig.~2(a--d)
			& Co$_3$O$_4$
			& 200
			& 21
			& $1.0\times10^{3}$
			& 0.20
			& 0.078 \\
			
			Fig.~2(h--k)
			& Amorph. C
			& 300
			& 19.68
			& $28.5\times10^{3}$
			& 0.25
			& 0.042 \\
			
			Fig.~3
			& CNT/TaTe$_2$
			& 80
			& 25
			& $30.5\times10^{3}$
			& 0.316
			& 0.029 \\
			
			Fig.~4(a)
			& Diffr. grating
			& 200
			& 30
			& $3.4\times10^{3}$
			& 0.727
			& 0.052 \\
			
			Fig.~4(b)
			& Gd$_2$O$_3$
			& 60
			& 30
			& $1.2\times10^{3}$
			& 0.43
			& 0.034 \\
			
			Fig.~4(c)
			& CNT/TaTe$_2$
			& 80
			& 19.68
			& $30.5\times10^{3}$
			& 0.25
			& 0.042 \\
			
			Fig.~5
			& Virus-like particle
			& 200
			& 30.6
			& 32
			& 11
			& 0.018 \\
			
			Fig.~6
			& Mitochondrion
			& 300
			& 7.0
			& 14
			& 28.7
			& 0.007 \\
			\bottomrule
		\end{tabular}
	\end{table}

	\subsection*{Viral-like particle preparation and grid preparation}
	
	Virus-like particles used in the experiment represent immature equine
	infectious anemia virus (EIAV) that, under chosen \textit{in~vitro}
	conditions, assemble into a mixture of spheres and tubes. The viral
	sample was prepared according to the protocol described in \cite{dick2020structures}.
	
	The CA--NC domains of EIAV Gag were cloned into a \texttt{pET28} vector
	in ORF with an N-terminal His$_6$--SUMO tag using a standard
	restriction--ligation molecular cloning method. The protein was
	expressed in \textit{E.~coli} BL21(DE3) cells. All purification steps
	were carried out at \(4^\circ\mathrm{C}\) or on ice. Cell pellets were
	resuspended in buffer (20~mM Tris--HCl pH~8, 500~mM NaCl, 2~mM TCEP,
	5~\(\mu\)M ZnCl$_2$), disrupted by sonication, and clarified by
	centrifugation. Nucleic acid was removed by adding 0.03\% (v/v)
	polyethyleneimine, and the resulting precipitate was removed by
	centrifugation. The target protein was then enriched by precipitation
	with 20\% ammonium sulfate, and the pellet was resuspended in the same
	buffer as used for affinity chromatography (20~mM Tris--HCl pH~8,
	500~mM NaCl, 2~mM TCEP). The resuspended material was filtered (0.2~\(\mu\)m)
	and loaded onto a Ni$^{2+}$-affinity column, followed by elution with
	imidazole. The eluate was dialyzed overnight in the presence of ULP1
	protease and was reapplied on a second Ni$^{2+}$ column to remove the
	cleaved SUMO tag and the ULP1 protease. The final purified protein was
	flash-frozen in liquid nitrogen and stored at \(-80^\circ\mathrm{C}\) in
	storage buffer (20~mM Tris--HCl pH~8, 500~mM NaCl, 2~mM TCEP).
	
	The assembly reaction was carried out in a buffer consisting of 50~mM
	MES (pH~6.0), 100~mM NaCl, 2~mM TCEP, supplemented with 10~\(\mu\)M IP6
	and 10~\(\mu\)M GT50. Reactions were incubated overnight at
	\(4^\circ\mathrm{C}\) and remained stable for several weeks thereafter.
	Grids were prepared on the Leica GP2 using backside blotting under 90\%
	humidity at \(10^\circ\mathrm{C}\). Before vitrification, the sample was
	mixed with 10~nm gold fiducial beads at a 1:10 (beads:sample) ratio. For
	grid preparation, 2~\(\mu\)L of the mixture was applied to the reverse side
	of the grid, and an additional 2.5~\(\mu\)L was applied to the front.
	Blotting was performed from the back for 3.5~s.

	\subsection*{Runtime performance}
\begin{figure*}[htbp!]
		\includegraphics[width=\textwidth]{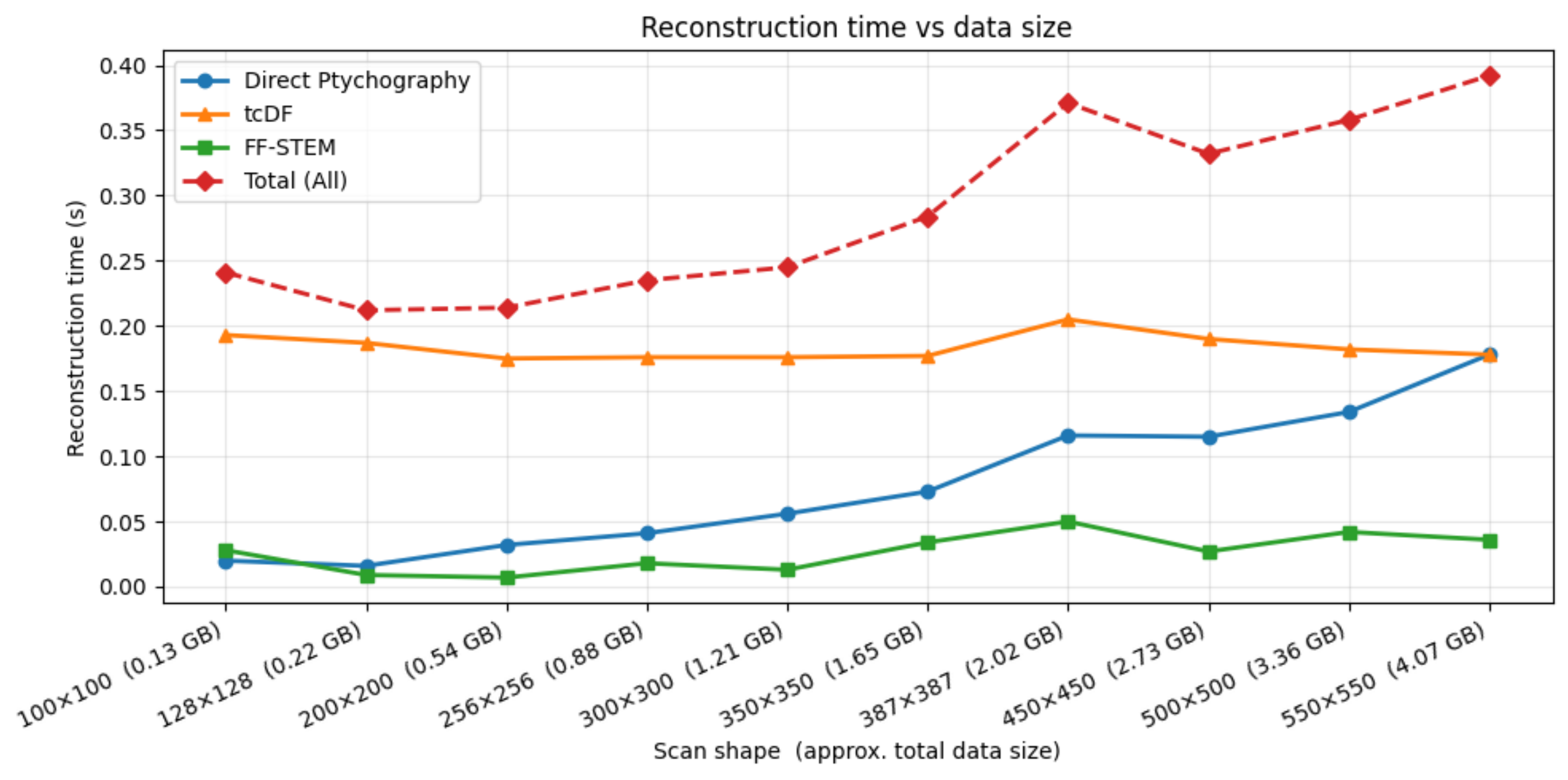}
		\caption{\label{fig:fig6} \textbf{Reconstruction time as a function of 4D-STEM data size} Measured reconstruction time for direct ptychography (blue), tcDF (orange), FF-STEM (green), and the total combined time (red). Each dataset corresponds to a cropped scan region of the full 4D-STEM acquisition. The approximate total data size (assuming 32-bit float) is indicated below the horizontal axis.}
	\end{figure*}

	We evaluate the runtime performance of our FF-STEM implementation by varying the number of scan positions used from the same experimental 4D-STEM dataset. As shown in Figure \ref{fig:fig6}, both the direct-ptychography and tcDF reconstructions exhibit nearly constant execution times across different scan sizes, reflecting their analytical, non-iterative formulation. The total runtime scales linearly with data volume, remaining below 0.5s even for the 550 x 550 x 85 x 85 dataset. This performance demonstrates the high computational efficiency of our CUDA-accelerated implementation and confirms that near-real-time reconstruction and fusion of megapixel-scale 4D-STEM data are feasible on a modern graphics processing unit. 
	The combination of accuracy, multi-frequency contrast, and sub-second runtime positions the FF-STEM method as a practical framework for live imaging and feedback during experiments. Due to its linear operation, it is compatible with live reconstructions from multi-segment detectors or streamed data from pixelated detectors, like OBF-STEM \cite{ooe2025dose} and various live ptychography implementations \cite{lalandec2025benchmarking,bangun2023wigner}.
	
	\section*{Availability of Data and Materials}
	The experimental data used in this study have been deposited at Zenodo under \href{https://doi.org/10.5281/zenodo.18008}{this link}.
	\section*{Author Contribution Statements}
	S.Y., P.P. and G.V. implemented FF-STEM algorithms. 
	S.Y. performed experimental reconstructions and prepared the initial manuscript.
	G.V. and P.P. derived and discussed SSNR and fusion weights.
	S.Y., M.W and E.S performed diffraction grating experiments.
	D.S. and P.P. performed cryo-4D-STEM experiment with VLPs.
	N.P. and S.S. performed 4D-STEM simulations and reconstructions of simulated data.
	S.K., R.E., S.V. performed Timepix4 experiments with $Gd_2O_3$. 
	F.S. and D.C. prepared frozen-hydrated virus-like particles and trained P.P in cryo-EM.
	B.Z. and X.Y. prepared $Gd_2O_3$ nanohelices.
	P.P. conceived the study.
	All authors read and approved the final version of the manuscript.
	
	\section*{Acknowledgements}
	We thank Tadahiro Yokosawa for support and discussions during the experiments.
	
	\section*{Financial Support}
	This project has received funding from the European Research Council (ERC) under the European Union’s Horizon 2020 research and innovation programme (Project HyperScaleEM, Grant agreement No. 101164581)
	\section*{Conflicts of Interest}
	The authors declare there are no conflicts of interest
	
	\selectlanguage{english}
	
	\bibliography{ref.bib%
	}
	
	\section*{Supplementary material}
	\subsubsection*{Derivative of \(\Gamma\) w.r.t. an aberration coefficient}
	
	Write \(A_0=A(\mathbf{k})\), \(A_\pm=A(\mathbf{k}\pm\mathbf{q})\) and \(\chi_0=\chi(\mathbf{k})\), \(\chi_\pm=\chi(\mathbf{k}\pm\mathbf{q})\). Using
	\[
	\frac{\partial A}{\partial \chi} = -iA,\quad
	\frac{\partial \overline{A}}{\partial \chi} = +i\overline{A},\quad
	\frac{\partial A}{\partial a_j} = -iA\,\frac{\partial\chi}{\partial a_j},\quad
	\frac{\partial \overline{A}}{\partial a_j} = +i\overline{A}\,\frac{\partial\chi}{\partial a_j},
	\]
	(neglecting \(\partial H/\partial a_j\) for a top-hat \(H\)), the product rule gives
	\begin{equation}
		\frac{\partial \Gamma}{\partial a_j}
		= i \left[
		\,\overline{A_0}A_-\,\Big(\tfrac{\partial\chi_0}{\partial a_j}-\tfrac{\partial\chi_-}{\partial a_j}\Big)
		+ A_0\overline{A_+}\,\Big(\tfrac{\partial\chi_0}{\partial a_j}-\tfrac{\partial\chi_+}{\partial a_j}\Big)
		\right].
	\end{equation}
	Equivalently, with \(C_-=\overline{A_0}A_-\) and \(C_+=A_0\overline{A_+}\),
	\begin{equation}
		\frac{\partial \Gamma}{\partial a_j}
		= i \left[
		C_1\left(\partial\chi_0-\partial\chi_-\right)
		+ C_2\left(\partial\chi_0-\partial\chi_+\right)
		\right].
	\end{equation}
	\subsection*{Isotropic TV loss and its gradient (2-D, \(\varepsilon\)-smoothed)}
	
	Let \(x\in\mathbb{R}^{H\times W}\) be the reconstructed phase image. We define forward differences
	\[
	\nabla_x x_{i,j}=x_{i,j+1}-x_{i,j},\qquad
	\nabla_y x_{i,j}=x_{i+1,j}-x_{i,j},
	\]
	with out-of-bounds forward diffs set to zero (Neumann boundary).
	
	The isotropic TV with \(\varepsilon>0\) smoothing is
	\[
	\mathcal{L}_{\text{TV}}(x)
	=\sum_{i=0}^{H-1}\sum_{j=0}^{W-1}
	\sqrt{(\nabla_x x_{i,j})^2+(\nabla_y x_{i,j})^2+\varepsilon^2}.
	\]
	Let
	\[
	s_{i,j}=\sqrt{(\nabla_x x_{i,j})^2+(\nabla_y x_{i,j})^2+\varepsilon^2},\qquad
	p_{i,j}=\frac{\nabla_x x_{i,j}}{s_{i,j}},\quad
	q_{i,j}=\frac{\nabla_y x_{i,j}}{s_{i,j}}.
	\]
	Then the gradient is the negative discrete divergence of \((p,q)\):
	\[
	\boxed{\displaystyle
		\frac{\partial \mathcal{L}_{\text{TV}}}{\partial x_{i,j}}
		= -\Big[(p_{i,j}-p_{i,j-1}) + (q_{i,j}-q_{i-1,j})\Big],}
	\]
	with \(p_{i,-1}=0\), \(q_{-1,j}=0\), and \(p,q\) zero wherever the corresponding forward difference is out-of-bounds.
	
	For interior pixels \(1\le i\le H-2,\;1\le j\le W-2\),
	\[
	\frac{\partial \mathcal{L}_{\text{TV}}}{\partial u_{i,j}}
	= -\left[
	\frac{x_{i,j+1}-x_{i,j}}{s_{i,j}}
	-\frac{x_{i,j}-x_{i,j-1}}{s_{i,j-1}}
	+\frac{x_{i+1,j}-x_{i,j}}{s_{i,j}}
	-\frac{x_{i,j}-x_{i-1,j}}{s_{i-1,j}}
	\right].
	\]
	We use the TotalVariation implementation in the pytorch package.
	\subsection*{Aperture autocorrelation}
	For the circular top-hat aperture in \eqref{eq:probe_forming_aperture_eq},
	the aperture autocorrelation depends only on $q = |\bm q|$ and can be
	written analytically as
	\begin{equation}
		[A \star A](q) =
		\begin{cases}
			2 k_0^2 \cos^{-1}\!\left( \dfrac{q}{2k_0} \right)
			- \dfrac{q}{2}\sqrt{4k_0^2 - q^2},
			& 0 \le q < 2k_0,\\[1.0em]
			0, & q \ge 2k_0.
		\end{cases}
		\label{eq:aperture_autocorr_analytic}
	\end{equation}
	Up to an overall normalization factor, \eqref{eq:aperture_autocorr_analytic}
	is the imaginary part of the ideal STEM CTF in \eqref{eq:aperture_autocorrelation_eq}.
	If the aperture $A(\bm k)$ is normalized such that
	$\int A(\bm k)\,d\bm k = 1$, the expression above should be multiplied
	by the factor $1/(\pi k_0^2)^2$.
	\subsection*{Supplementary Information}
	\subsubsection*{Simulated Depth sections}
	\begin{figure*}[htbp!]
		\includegraphics[width=\textwidth]{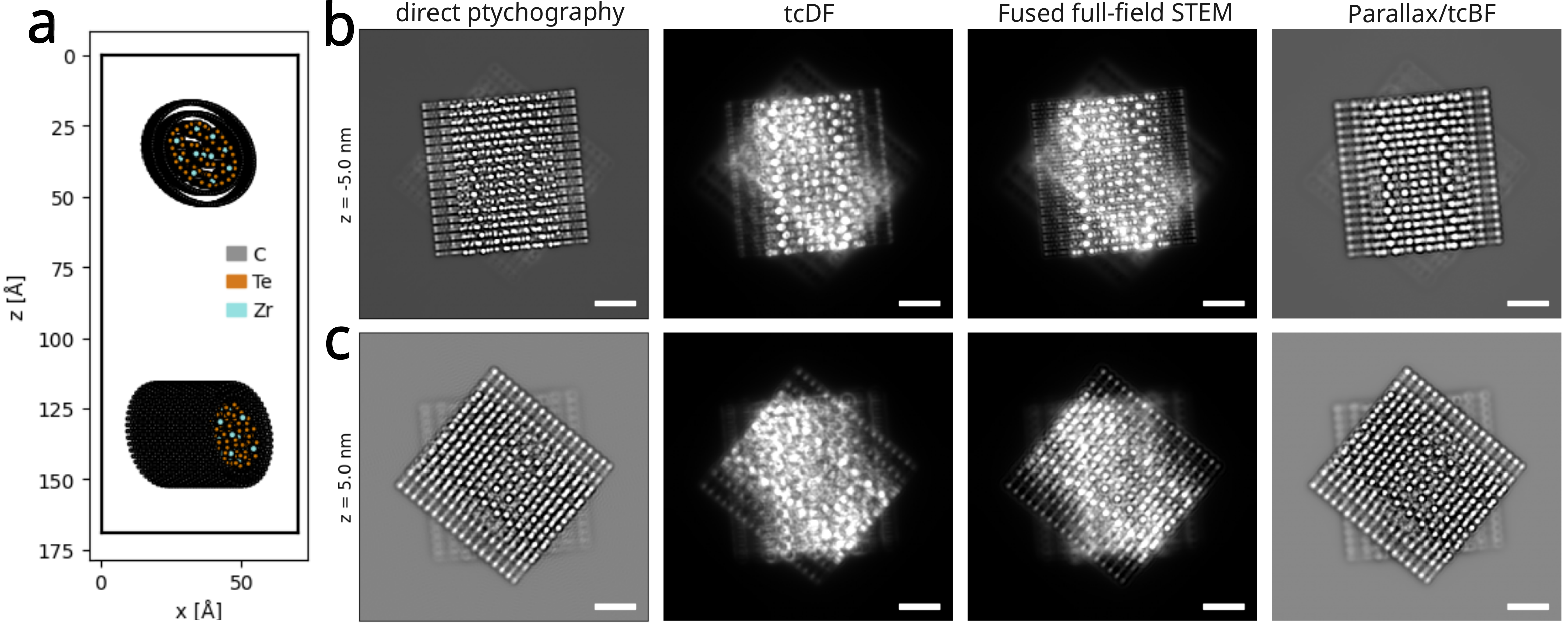}
		\caption{\label{fig:fig8} \textbf{Simulated depth-sectioning of double-wall carbon nanotubes (DW-CNTs) encapsulating a Zr-Te core using FF-STEM reconstruction.} (a) Atomic model of the simulated sample consisting of two identical DW-CNTs containing a complex Zr-Te sandwich structure, separated by 10nm along the beam direction(z). (b) Reconstructed images from direct ptychography, tcDF, FF-STEM, and parallax methods, with the focal plane set to the upper nanotube (z=-5nm). (c) Corresponding reconstructions with the focal plane shifted to the lower nanotube (z=+5nm). The scale bar is 5\SI{}{\angstrom}.}
	\end{figure*}
	
	The depth-sectioning capability of FF-STEM is demonstrated on a simulation cell containing two identical DW-CNT encapsulating a Zr-Te core \cite{pelz2023solving}, separated by \SI{10}{\nano\meter} along the optical axis. The simulated dataset was reconstructed using direct ptychography, tcDF, FF-STEM and parallax methods, with the defocus adjusted to sequentially focus on the top (z=-5nm) and bottom (z=+5nm)nanotube, as shown in Figure \ref{fig:fig8}(b, c). 
	
	Under infinite-dose conditions, both FF-STEM and parallax reconstructions exhibit excellent lattice resolution and accurate depth localization, clearly distinguishing the two axial planes. the direct ptychography results show similarly sharp atomic features but, as expected, reduced low-frequency contrast, whereas the tcDF images appear blurred and lack high-frequency detail. The FF-STEM reconstruction combines the complementary strengths of the two input channels, maintaining phase fidelity and improved intensity uniformity across focal panes. These results confirm that the FF-STEM framework naturally extends to depth-sectioned phase imaging, preserving atomic resolution while discriminating features separated by several nanometers along the beam direction.
	
	\begin{figure}
		\textbf{Table of Contents}\\
		\medskip
		\includegraphics{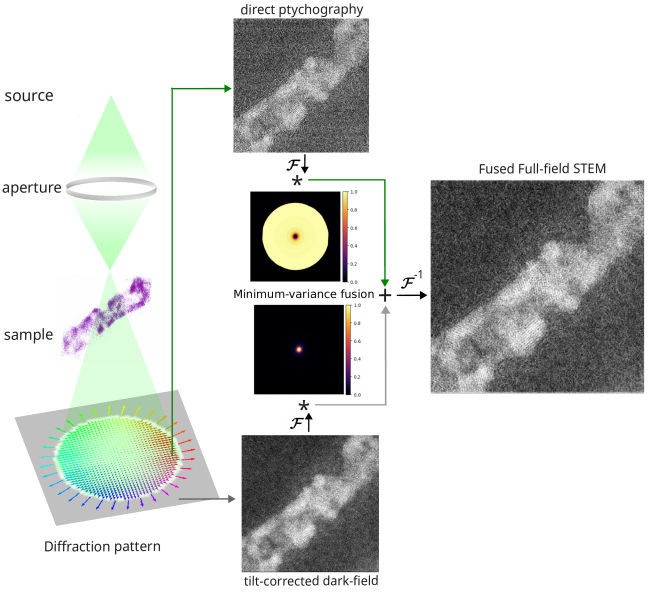}
		\medskip
		\caption*{Fused Full-Field STEM (FF-STEM) combines complementary bright-field and dark-field scattering channels in 4D-STEM using an optimal signal-to-noise-based fusion strategy. By uniting phase-sensitive high-resolution information with robust low-frequency contrast, FF-STEM enables gap-free, dose-efficient and high-contrast imaging from a single dataset}
	\end{figure}
	
\end{document}